\documentclass[preprint,prd,aps,showpacs,showkeys,nofootinbib]{revtex4}
\usepackage{graphicx}
\usepackage{dcolumn}
\usepackage{bm}
\usepackage{color}
\usepackage[dvipsnames]{xcolor}
\usepackage{amssymb}

\definecolor{light-gray}{gray}{0.78}
\definecolor{mid-gray}{gray}{0.55}
\definecolor{dark-gray}{gray}{0.32}

\begin{document}
\title{Lepton flavor violating decays $l_j\rightarrow l_i\gamma$, $l_j \rightarrow 3l_i$ and $\mu\rightarrow e+ q\bar q$ in the N-B-LSSM}
\author{Rong-Zhi Sun$^{1,2,3}$, Shu-Min Zhao$^{1,2,3}$\footnote{zhaosm@hbu.edu.cn}, Ming-Yue Liu$^{1,2,3}$, Xing-Yu Han$^{1,2,3}$, Song Gao$^{1,2,3}$, Xing-Xing Dong$^{1,2,3,4}$\footnote{dongxx@hbu.edu.cn}}
\affiliation{$^1$ Department of Physics, Hebei University, Baoding 071002, China}
\affiliation{$^2$ Hebei Key Laboratory of High-precision Computation and Application of Quantum Field Theory, Baoding, 071002, China}
\affiliation{$^3$ Hebei Research Center of the Basic Discipline for Computational Physics, Baoding, 071002, China}
\affiliation{$^4$ Departamento de Fisica and CFTP, Instituto Superior T$\acute{e}$cnico, Universidade de Lisboa,
Av.Rovisco Pais 1,1049-001 Lisboa, Portugal}
\date{\today}

\begin{abstract}

The N-B-LSSM is an extension of the minimal supersymmetric standard model (MSSM) with the addition of three singlet new Higgs superfields and right-handed neutrinos, whose local gauge group is $SU(3)_C\times SU(2)_L\times U(1)_Y\times U(1)_{B-L}$. In the N-B-LSSM, we study lepton flavor violating decays $l_j\rightarrow l_i\gamma$, $l_j \rightarrow 3l_i$ and $\mu\rightarrow e+ q\bar q$ $(j=\tau,\mu,~i=\mu,e$ and $i\neq j)$. Based on the current experimental limitations, we carry out detailed parameter scanning and numerical calculations to analyse the effects of different sensitive parameters on lepton flavor violation (LFV) in the N-B-LSSM. The numerical results show that the non-diagonal elements involving the initial and final leptons are main sensitive parameters and LFV sources. This work can provide a strong basis for exploring new physics (NP) beyond the Standard Model (SM).

\end{abstract}

\maketitle
\section{Introduction}

Over the past few decades, the successful observation of neutrino oscillations \cite{neutrino1,neutrino2,neutrino3} has not only confirmed that neutrinos possess non-zero masses but also revealed significant lepton flavor mixing \cite{neutrinoN1,neutrinoN2,neutrinoN3,neutrinoN4}, thereby challenging the fundamental assumption of lepton flavor conservation in the SM. Although the SM has achieved remarkable success in describing strong, weak, and electromagnetic interactions, the presence of the Glashow-Iliopoulos-Maiani (GIM) mechanism leads to an extreme suppression of the predicted branching ratios for LFV decays \cite{p1}. For instance, the SM prediction for the branching ratio of $\mu\rightarrow e\gamma$ is as low as about $10^{-55}$, far below the current experimental sensitivity. Consequently, any observation of LFV signals would provide compelling evidence for new physics (NP) beyond the SM. The paper summarizes the latest experimental results, including the upper limits on the LFV branching ratios of $l_j\rightarrow l_i\gamma$ and $l_j \rightarrow 3l_i$ at 90\% confidence level (CL) \cite{pdg}, as well as the current sensitivities of $\mu -e$ conversion rates
in different nuclei \cite{experiment1,experiment2,experiment3}:
\begin{eqnarray}
&&{\rm Br}(\mu\rightarrow e\gamma)<4.2\times10^{-13},~{\rm Br}(\mu\rightarrow 3e)<1.0\times10^{-12},~{\rm CR}(\mu\rightarrow e:{\rm Au})<7.0\times10^{-13},
\nonumber\\&&{\rm Br}(\tau\rightarrow \mu\gamma)<4.2\times10^{-8},~{\rm Br}(\tau\rightarrow 3\mu)<2.1\times10^{-8},~{\rm CR}(\mu\rightarrow e:{\rm Ti})<4.3\times10^{-12},
\nonumber\\&&{\rm Br}(\tau\rightarrow e\gamma)<3.3\times10^{-8},~{\rm Br}(\tau\rightarrow 3e)<2.7\times10^{-8},~{\rm CR}(\mu\rightarrow e:{\rm Pb})<4.6\times10^{-11}.
\end{eqnarray}

To address the shortcomings of the SM in accounting for LFV phenomena, supersymmetry (SUSY) models have been widely investigated as extensions of the SM, with the MSSM being the most prominent example. However, the MSSM still has limitations in solving core challenges such as the $\mu$ problem \cite{mu} and the zero mass neutrino \cite{neutrino4}, which motivates the development of more comprehensive frameworks.

To further refine MSSM and overcome these issues, next to the minimal supersymmetric extension of the SM with local $B-L$ gauge symmetry (N-B-LSSM) has been proposed \cite{han}. The key feature of this model is the incorporation of an additional $U(1)_{B-L}$ gauge symmetry, which extends the MSSM gauge group to $SU(3)_C\times SU(2)_L \times U(1)_Y\times U(1)_{B-L}$, where $B$ denotes the baryon number and $L$ represents the lepton number. This extension allows for a more natural explanation of neutrino masses and provides a framework for lepton and baryon number violation. One of the major innovations of N-B-LSSM is the introduction of right-handed neutrino superfields and three Higgs singlet superfields. This not only provides a natural mechanism for generating small neutrino masses through the seesaw mechanism but also offers an effective solution to the $\mu$ problem, which MSSM fails to resolve. Specifically, in N-B-LSSM, the Higgs singlet $\hat{S}$, with a non-zero vacuum expectation value (VEV) of $\frac{v_S}{\sqrt{2}}$, couples to the up-type and down-type Higgs doublets $H_u$ and $H_d$, generating the interaction term $\lambda\hat{S}\hat{H}_u\hat{H}_d$. This term replaces the traditional $\mu$ term in MSSM, resulting in an effective mass term $\lambda\frac{v_S}{\sqrt{2}}\hat{H}_u\hat{H}_d$, thus solving the $\mu$ problem without requiring additional fine-tuning. The extended Higgs sector leads to a 5$\times$5 neutral CP-even Higgs mass matrix, which provides an explanation for the 125.20 $\pm$ 0.11 GeV Higgs mass. In the N-B-LSSM, the right-handed neutrinos, singlet Higgs fields, and the additional superfields effectively mitigate the little hierarchy problem in the MSSM through their higher VEVs. Assuming that the VEVs of these superfields are located at higher scales, the new particles' contributions to the model are effectively suppressed.

In supersymmetric extensions of the SM, both the $\mu\nu$SSM \cite{ZHB1,ZHB2} and the $\nu_R$MSSM \cite{nuRMSSM} address the neutrino mass problem in the SM by introducing right-handed neutrinos ($\nu_R$). On this basis, they also enhance the amplitudes of LFV decays. In the supersymmetric standard
model with right-handed neutrino supermultiplets, the authors investigate various LFV processes in detail \cite{ljlig}. A minimal supersymmetric extension of the SM with local gauged $B$ and $L$(BLMSSM) is first proposed by the author \cite{BLMSSM3,BLMSSM4}. The $\mu\rightarrow e$ conversion is investigated within the BLMSSM framework \cite{GT}. In Ref. \cite{han}, N-B-LSSM is proposed for the first time, which is the model adopted in this paper. In previous work, some two loop contribution to muon anomalous MDM in the N-B-LSSM is studied. This paper investigates the LFV processes ($\mu\rightarrow e\gamma$, $\tau\rightarrow e\gamma$, $\tau\rightarrow\mu\gamma$; $\mu\rightarrow3e$, $\tau\rightarrow3e$, $\tau\rightarrow3\mu$; $\mu\rightarrow e+ q\bar q$) in the framework of N-B-LSSM. We derive the corresponding Feynman diagrams, and the Feynman amplitudes, decay widths, branching ratios are analyzed numerically. Under the constraints of the latest experimental limits, sensitive and insensitive parameters are identified through one dimensional plots or scatter plots.

The structure of this paper is as follows. In Sec.II, we briefly summarize the main content of the N-B-LSSM. In Sec.III, we derive and present the analytical expressions for the branching ratios for $l_j \rightarrow l_i\gamma$ and $l_j \rightarrow 3l_i$, as well as the conversion rate for $\mu\rightarrow e+ q\bar q$ within the N-B-LSSM framework. The input parameters and numerical analysis are detailed in Sec.IV. Our conclusions are presented in Sec.V. Finally, some mass matrices and couplings are collected in the Appendix \ref{A1}.

\section{The main content of N-B-LSSM}

N-B-LSSM is an extension of MSSM, introducing an additional local gauge symmetry {$U(1)_{B-L}$}. The local gauge group of this model is $SU(3)_C\otimes SU(2)_L\otimes U(1)_Y\otimes U(1)_{B-L}$. Compared to MSSM, N-B-LSSM includes new superfields, such as right-handed neutrinos $\hat{\nu_i}$ and three Higgs singlets $\hat{\chi}_1,~\hat{\chi}_2,~\hat{S}$. The {$U(1)_{B-L}$} symmetry is spontaneously broken by the VEVs of $\chi_1$ and $\chi_2$, which also generate the large Majorana masses for right-handed neutrinos. Through the seesaw mechanism, light neutrinos obtain tiny masses at the tree level. The neutral CP-even parts of $H_u$, $H_d$, $\chi_1$, $\chi_2$ and $S$ mix with each other, forming a $5\times5$ mass squared matrix. The lightest mass eigenvalue within this matrix corresponds to the lightest CP-even Higgs. At the tree level, the theoretical Higgs mass generally does not match the experimentally observed value of 125.20 $\pm$ 0.11 GeV \cite{LCTHiggs1,LCTHiggs2}. To resolve this discrepancy, loop corrections must be included. As for sneutrinos, they are further classified into CP-even sneutrinos and CP-odd sneutrinos, and their mass squared matrices are expanded to $6\times6$ due to the inclusion of right-handed sneutrinos and their interactions.

The superpotential of N-B-LSSM is :
\begin{eqnarray}
&&W=-Y_d\hat{d}\hat{q}\hat{H}_d-Y_e\hat{e}\hat{l}\hat{H}_d-\lambda_2\hat{S}\hat{\chi}_1\hat{\chi}_2+\lambda\hat{S}\hat{H}_u\hat{H}_d+\frac{\kappa}{3}\hat{S}\hat{S}\hat{S}+Y_u\hat{u}\hat{q}\hat{H}_u+Y_{\chi}\hat{\nu}\hat{\chi}_1\hat{\nu}
\nonumber\\&&~~~~~~~+Y_\nu\hat{\nu}\hat{l}\hat{H}_u.
\end{eqnarray}

In the superpotential for this model, the Yukawa couplings are denoted by $Y_{u,d,e,\nu,\chi}$. While $\lambda$, $\lambda_2$ and $\kappa$ represent dimensionless couplings. The fields $\hat{\chi}_1,~\hat{\chi}_2,~\hat{S}$ are Higgs singlets. It is important to note that the term $Y^\prime_\nu\hat{\nu}\hat{l}\hat{S}$ is not allowed, as the sum of $U(1)_Y$ charges of $\hat{\nu},\hat{l},\hat{S}$ does not satisfy the necessary charge neutrality condition.

We show the concrete forms of the two Higgs doublets and three Higgs singlets

\begin{eqnarray}
&&\hspace{1cm}H_{u}=\left(\begin{array}{c}H_{u}^+\\{1\over\sqrt{2}}\Big(v_{u}+H_{u}^0+iP_{u}^0\Big)\end{array}\right),
~~~~~~
H_{d}=\left(\begin{array}{c}{1\over\sqrt{2}}\Big(v_{d}+H_{d}^0+iP_{d}^0\Big)\\H_{d}^-\end{array}\right),
\nonumber\\&&\chi_1={1\over\sqrt{2}}\Big(v_{\eta}+\phi_{1}^0+iP_{1}^0\Big),~~~
\chi_2={1\over\sqrt{2}}\Big(v_{\bar{\eta}}+\phi_{2}^0+iP_{2}^0\Big),~~
S={1\over\sqrt{2}}\Big(v_{S}+\phi_{S}^0+iP_{S}^0\Big).
\end{eqnarray}

The VEVs of the Higgs superfields $H_u$, $H_d$, $\chi_1$, $\chi_2$ and $S$ are presented by
$v_u,~v_d,~v_\eta$,~ $v_{\bar\eta}$ and $v_S$ respectively. Two angles are defined as $\tan\beta=v_u/v_d$ and $\tan\beta_\eta=v_{\bar{\eta}}/v_{\eta}$. The definitions of ${\widetilde{\nu}}_{L}$ and ${\widetilde{\nu}}_{R}$ are:
\begin{eqnarray}
&&\widetilde{\nu}_{L}= \frac{1}{\sqrt{2} } {\phi}_{L}+\frac{i}{\sqrt{2} } {\sigma}_{L},~~~~~~~~~~~~~~
\widetilde{\nu}_{R}= \frac{1}{\sqrt{2} } {\phi}_{R}+\frac{i}{\sqrt{2} } {\sigma}_{R}.
\end{eqnarray}

The soft SUSY breaking terms of N-B-LSSM are shown as:
\begin{eqnarray}
&&\mathcal{L}_{soft}=\mathcal{L}_{soft}^{MSSM}-\frac{T_\kappa}{3}S^3+\epsilon_{ij}T_{\lambda}SH_d^iH_u^j+T_{2}S\chi_1\chi_2\nonumber\\&&
-T_{\chi,ik}\chi_1\tilde{\nu}_{R,i}^{*}\tilde{\nu}_{R,k}^{*}
+\epsilon_{ij}T_{\nu,ij}H_u^i\tilde{\nu}_{R,i}^{*}\tilde{e}_{L,j}-m_{\eta}^2|\chi_1|^2-m_{\bar{\eta}}^2|\chi_2|^2\nonumber\\&&-m_S^2|S|^2-m_{\nu,ij}^2\tilde{\nu}_{R,i}^{*}\tilde{\nu}_{R,j}
-\frac{1}{2}(2M_{BB^\prime}\lambda_{\tilde{B}}\tilde{B^\prime}+M_{BL}\tilde{B^\prime}^2)+h.c~~.\label{L}
\end{eqnarray}

In the Eq.(\ref{L}),~$\mathcal{L}_{soft}^{MSSM}$ represents the soft supersymmetry-breaking terms of MSSM. The parameters $T_{\kappa}$, $T_{\lambda}$, $T_2$, $T_{\chi}$ and $T_{\nu}$ are trilinear coupling coefficients.

\begin{table}[h]
\caption{ The superfields in N-B-LSSM}
\begin{tabular}{|c|c|c|c|c|}
\hline
Superfields & $SU(3)_C$ & $SU(2)_L$ & $U(1)_Y$ & $U(1)_{B-L}$ \\
\hline
$\hat{q}$ & 3 & 2 & 1/6 & 1/6  \\
\hline
$\hat{l}$ & 1 & 2 & -1/2 & -1/2  \\
\hline
$\hat{H}_d$ & 1 & 2 & -1/2 & 0 \\
\hline
$\hat{H}_u$ & 1 & 2 & 1/2 & 0 \\
\hline
$\hat{d}$ & $\bar{3}$ & 1 & 1/3 & -1/6  \\
\hline
$\hat{u}$ & $\bar{3}$ & 1 & -2/3 & -1/6 \\
\hline
$\hat{e}$ & 1 & 1 & 1 & 1/2 \\
\hline
$\hat{\nu}$ & 1 & 1 & 0 & 1/2 \\
\hline
$\hat{\chi}_1$ & 1 & 1 & 0 & -1 \\
\hline
$\hat{\chi}_2$ & 1 & 1 & 0 & 1\\
\hline
$\hat{S}$ & 1 & 1 & 0 & 0 \\
\hline
\end{tabular}
\label{biao1}
\end{table}
The particle content and charge distribution for N-B-LSSM are shown in the Table \ref {biao1}. In the chiral superfields, $\hat H_u = \Big( {\hat H_u^ + ,\hat H_u^0} \Big)$    and $\hat H_d = \Big( {\hat H_d^0,\hat H_d^ - } \Big)$ represent the MSSM-like Higgs doublet superfields. The superfields $\hat q $ and $\hat l $ are the doublets of quarks and leptons. The singlet superfields include $\hat u$, $\hat d$, $\hat e$ and $\hat{\nu}$, which correspond to the up-type quark, down-type quark, charged lepton and neutrino superfields.

$Y^Y$ signifies the $U(1)_Y$ charge, while $Y^{B-L}$ denotes the {$U(1)_{B-L}$} charge. The two Abelian groups, $U(1)_Y$ and  {$U(1)_{B-L}$}, within the N-B-LSSM, produce a new effect: the gauge kinetic mixing. This effect can be induced trough RGEs, even when it starts from a zero value at $M_{GUT}$. Since both Abelian gauge groups remain unbroken, a basis transformation is permissible through the rotation matrix $R$ ($R^T R=1$) \cite{UMSSM5,B-L1,B-L2,gaugemass}.

The form of the covariant derivatives of the N-B-LSSM  can be written as:
{\begin{eqnarray}
&&D_\mu=\partial_\mu-i\left(\begin{array}{cc}Y,&B-L\end{array}\right)
\left(\begin{array}{cc}g_{Y},&g{'}_{{YB}}\\g{'}_{{BY}},&g{'}_{{B-L}}\end{array}\right)
\left(\begin{array}{c}B_{\mu}^{\prime Y} \\ B_{\mu}^{\prime BL}\end{array}\right)\;.
\end{eqnarray}}

$B_\mu^{\prime Y}$ and $B_\mu^{\prime BL}$ denote the gauge fields pertaining to the $U(1)_Y$ and  {$U(1)_{B-L}$} respectively. Under the condition that the aforementioned Abelian gauge symmetry groups remain unbroken, the transformation of the basis is carried out through the application of a rotation matrix $R$.

{\begin{eqnarray}
&&D_\mu=\partial_\mu-i\left(\begin{array}{cc}Y^Y,&Y^{B-L}\end{array}\right)
\left(\begin{array}{cc}g_{Y},&g{'}_{{YB}}\\g{'}_{{BY}},&g{'}_{{B-L}}\end{array}\right)
R^TR
\left(\begin{array}{c}B_{\mu}^{\prime Y} \\ B_{\mu}^{\prime BL}\end{array}\right)\;,
\end{eqnarray}}
with the redefinitions
\begin{eqnarray}
&&\left(\begin{array}{cc}g_{Y},&g{'}_{{YB}}\\g{'}_{{BY}},&g{'}_{{B-L}}\end{array}\right)
R^T=\left(\begin{array}{cc}g_{1},&g_{{YB}}\\0,&g_{{B}}\end{array}\right)~~~~\text{and}~~~~~
R\left(\begin{array}{c}B_{\mu}^{\prime Y} \\ B_{\mu}^{\prime BL}\end{array}\right)
=\left(\begin{array}{c}B_{\mu}^{Y} \\ B_{\mu}^{BL}\end{array}\right)\;,
\end{eqnarray}

Then the covariant derivatives of this model can be changed as
{\begin{eqnarray}
&&D_\mu=\partial_\mu-i\left(\begin{array}{cc}Y^Y,&Y^{B-L}\end{array}\right)
\left(\begin{array}{cc}g_1,&g_{{YB}}\\0,&g_B\end{array}\right)
\left(\begin{array}{c}B_{\mu}^{Y} \\ B_{\mu}^{BL}\end{array}\right)\;.
\end{eqnarray}}

Within the framework of this model, $g_B$ is defined as the gauge coupling constant corresponding to the $U(1)_{B-L}$ group, and $g_{YB}$ denotes the mixing gauge coupling constant of $U(1)_{B-L}$ group and $U(1)_Y$ group, the gauge bosons denoted by $B^{Y}_\mu,~B^{{BL}}_\mu$ and $V^3_\mu$ intermingle at the tree level. The corresponding mass matrix is defined in the basis $(B^{Y}_\mu, B^{{BL}}_\mu, V^3_\mu)$:
\begin{eqnarray}
&&\left(\begin{array}{*{20}{c}}
\frac{1}{8}g_{1}^2 v^2 &~~~ -\frac{1}{8}g_{1}g_{2} v^2 & ~~~\frac{1}{8}g_{1}(g_{YB}+g_B) v^2 \\
-\frac{1}{8}g_{1}g_{2} v^2 &~~~ \frac{1}{8}g_{2}^2 v^2 & ~~~~-\frac{1}{8}g_{2}(g_{YB}+g_B) v^2\\
\frac{1}{8}g_{1}(g_{YB}+g_B) v^2 &~~~ -\frac{1}{8}g_{2}(g_{YB}+g_B) v^2 &~~~~ \frac{1}{8}(g_{YB}+g_B)^2 v^2+\frac{1}{8}g_{{B}}^2 \xi^2
\end{array}\right),\label{matrix}
\end{eqnarray}
with $v^2=v_u^2+v_d^2$ and $\xi^2=v_\eta^2+v_{\bar{\eta}}^2$. The mass matrix in Eq.(\ref{matrix}) is diagonalized using the Weinberg angle $\theta_{W}$ and the new mixing angles $\theta_{W}'$.
$\theta_{W}'$ is defined from the following formula:

\begin{eqnarray}
\sin^2\theta_{W}'\!=\!\frac{1}{2}\!-\!\frac{[(g_{{YB}}+g_{B})^2-g_{1}^2-g_{2}^2]v^2+
4g_{B}^2\xi^2}{2\sqrt{[(g_{{YB}}+g_{B})^2+g_{1}^2+g_{2}^2]^2v^4\!+\!8g_{B}^2[(g_{{YB}}+g_{B})^2\!-\!g_{1}^2\!-\!g_{2}^2]v^2\xi^2\!+\!16g_{B}^4\xi^4}}.
\end{eqnarray}

We deduce the eigenvalues of Eq.(\ref{matrix})
\begin{eqnarray}
&&m_\gamma^2=0,\nonumber\\
&&m_{Z,{Z^{'}}}^2=\frac{1}{8}\Big([g_{1}^2+g_2^2+(g_{{YB}}+g_{B})^2]v^2+4g_{B}^2\xi^2 \nonumber\\
&&\hspace{1.1cm}\mp\sqrt{[g_{1}^2+g_{2}^2+(g_{{YB}}+g_{B})^2]^2v^4\!+\!8[(g_{{YB}}+g_{B})^2\!-\!g_{1}^2\!-\!
g_{2}^2]g_{B}^2v^2\xi^2\!+\!16g_{B}^4\xi^4}\Big).
\end{eqnarray}

We show some used mass matrixes and couplings in the Appendix \ref{A1}.

\section{formulation}

In this section, the LFV processes  $l_j\rightarrow l_i\gamma,~l_j \rightarrow 3l_i~(j=\tau,\mu,~i=\mu,e$ and $i\neq j)$ and $\mu\rightarrow \emph{e}+\rm{q}\bar q$ are studied in the N-B-LSSM. The effective Lagrangian is affected by contributions from one-loop, triangle-type, penguin-type, self-energy and box-type diagrams. For convenience, these diagrams are analyzed in the generic form, which can simplify the work.

\subsection{$l_j\rightarrow l_i\gamma$}

\begin{figure}[h]
\setlength{\unitlength}{5.0mm}
\centering
\includegraphics[width=5.0in]{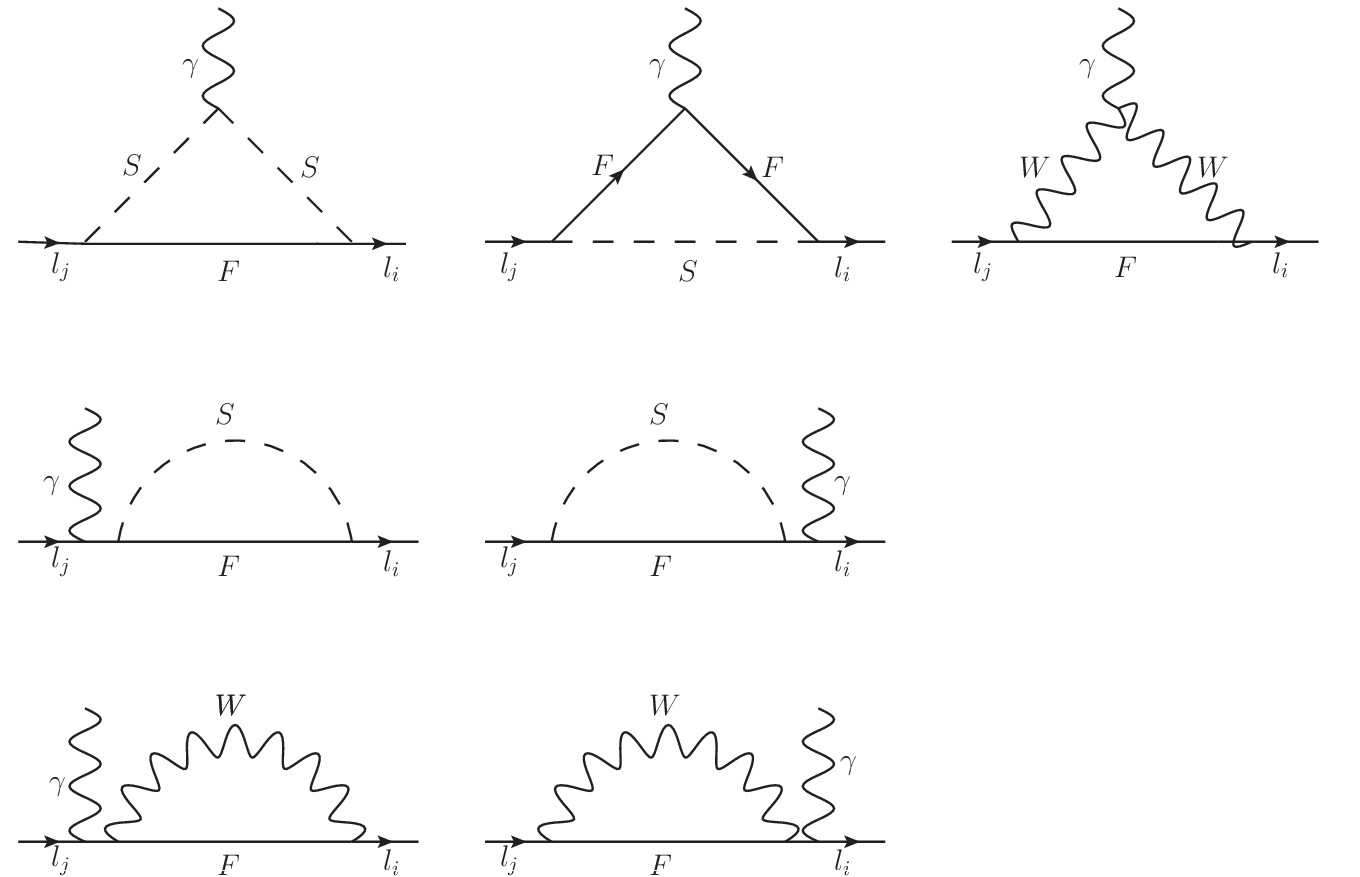}
\caption{ The one-loop diagrams for $l_j\rightarrow{l_i\gamma}$, with $F$ representing Dirac (Majorana) particles.}\label{N1}
\end{figure}

The relevant Feynman diagrams are shown in Fig.\ref{N1}. When the external leptons are all on shell, the amplitudes for $l_j\rightarrow l_i\gamma$ can be expressed in the following general form:

\begin{eqnarray}
&&\mathcal{M} = e{\epsilon ^\mu }{\bar u_i}(p + q)\Big[q^2{\gamma _\mu }(A_1^LP_L + A_1^RP_R)
\nonumber\\
&&\qquad + \: {m_{{l_j}}}i{\sigma _{\mu \nu }}{q^\nu }(A_2^LP_L + A_2^RP_R)\Big]{u_j}(p)\:,
\label{amplitude-gamma}
\end{eqnarray}
where $p$ denotes the injecting lepton momentum, $q$ represents the photon momentum, and $m_{{l_j}}$ is the mass of the charged lepton in the $j$th generation. The wave function for the external leptons are given by ${u_i}(p)$ and ${v_i}(p)$. The final Wilson coefficients $A_1^L,A_1^R,A_2^L,A_2^R$ are determined by summing the amplitudes of the corresponding diagrams.

The contributions from the virtual neutral fermion diagrams are denoted by $A^{L,R}_{\alpha}(n), \alpha=1,2$. The derived results are presented in the following form:

\begin{eqnarray}
&&A_{1}^{L}(n)=\sum_{\beta=1}^8\sum_{\alpha=1}^6\frac{1}{6{m_W^2}}C_R^{\tilde{L}_\alpha\chi^0_\beta\bar{l}_i}C_L^{\tilde{L}_\alpha^*l_j\bar{\chi}_\beta^0}{F_1}
(x_{\chi_\beta^0},x_{\tilde{L}_\alpha})\nonumber\\
&&~~~~~~~~+\sum_{\delta=1}^6\sum_{\vartheta=1}^2\frac{1}{6{m_W^2}}C_R^{H^{\pm}_\vartheta\nu_\delta\bar{l}_i}C_L^{H^{\pm *}_\vartheta l_j\bar{\nu}_\delta}{F_1}
(x_{\nu_\delta},x_{H^{\pm}_\vartheta}),\nonumber\\
&&A_{2}^{L}(n)=\sum_{\beta=1}^8\sum_{\alpha=1}^6\frac{{{m_{\chi^0_\beta}}}}{{{m_{{l_j}}}}{m_W^2}}C_L^{\tilde{L}_\alpha\chi^0_\beta\bar{l}_i}C_L^{\tilde{L}_\alpha^*l_j\bar{\chi}_\beta^0}{F_2}(x_{\chi_\beta^0},x_{\tilde{L}_\alpha})\nonumber\\
&&~~~~~~~~+\sum_{\delta=1}^6\sum_{\vartheta=1}^2\frac{{{m_{\nu_\delta}}}}{{{m_{{l_j}}}}{m_W^2}}C_L^{H^{\pm}_\vartheta\nu_\delta\bar{l}_i}C_L^{H^{\pm *}_\vartheta l_j\bar{\nu}_\delta}{F_2}
(x_{\nu_\delta},x_{H^{\pm}_\vartheta}),\nonumber\\
&&A_\alpha^{R}(n) = \left. {A_\alpha^{L}}(n) \right|{ _{L \leftrightarrow R}},~~~\alpha=1,2.\label{oneloopn}
\end{eqnarray}

Here, $x_i= {m_i^2}/{m_W^2}$ with $m_i$ denoting the mass of the corresponding particle, the one-loop functions $F_1(x,y)$ and $F_2(x,y)$ are compiled as follows:
\begin{eqnarray}
&&F_1(x,y)=\frac{1}{96\pi^2}\Big[\frac{7xy-11x^2-2y^2}{(x-y)^3}+\frac{6x^3(\ln{x}-\ln{y})}{(x-y)^4}\Big],\nonumber\\
&&F_2(x,y)=\frac{1}{32\pi^2}\Big[-\frac{x+y}{(x-y)^2}+\frac{2xy(\ln{x}-\ln{y})}{(x-y)^3}\Big].
\end{eqnarray}

The coefficients $A^{L,R}_{\alpha}(c),\alpha=1,2$ represent contributions from the virtual charged fermion diagrams and the expressions are:
\begin{eqnarray}
&&A^L_1(c)=\sum_{\rho=1}^2\sum_{\sigma=1}^6\frac{1}{6m^2_W}C^{\tilde{\nu}_\sigma\chi^{\pm}_\rho\bar{l}_i}_RC^{\tilde{\nu}_\sigma^*{l_j}\bar{\chi}^{\pm}_\rho}_L{F_3}(x_{\chi^{\pm}_\rho},x_{\tilde{\nu}_\sigma}),\nonumber\\
&&A^L_2(c)=\sum_{\rho=1}^2\sum_{\sigma=1}^6\frac{m_{\chi^{\pm}_\rho}}{m_{l_j}m^2_W}C^{\tilde{\nu}_\sigma\chi^{\pm}_\rho\bar{l}_i}_LC^{\tilde{\nu}_\sigma^*{l_j}\bar{\chi}^{\pm}_\rho}_L{F_4}(x_{\chi^{\pm}_\rho},x_{\tilde{\nu}_\sigma}),\nonumber\\
&&A^R_{\alpha}(c)=A^L_{\alpha}(c)|_{L\leftrightarrow{R}},~~~\alpha=1,2.\label{oneloopc}
\end{eqnarray}
with
\begin{eqnarray}
&&F_3(x,y)=\frac{1}{96\pi^2}\Big[\frac{29xy-7x^2-16y^2}{(x-y)^3}-\frac{6y^2(3x-2y) (\ln{x}-\ln{y})}{(x-y)^4}\Big],\nonumber\\
&&F_4(x,y)=\frac{1}{96\pi^2}\Big[-\frac{-17xy+x^2+10y^2}{(x-y)^3}-\frac{6y\left(x^2+xy-y^2\right)(\ln{x}-\ln{y})}{(x-y)^4}\Big].
\end{eqnarray}

In this paper, $\tilde{\nu}$ encompasses both $\tilde{\nu}^R$ and $\tilde{\nu}^I$.

The mixing of three light neutrinos with three heavy neutrinos introduces corrections to the LFV processes $l_j\rightarrow l_i\gamma$ through virtual $W$ diagrams. The corresponding coefficients are denoted as
$A_\alpha^{L,R}(W),\alpha=1,2$.
\begin{eqnarray}
&&A^L_1(W)=\sum_{\delta=1}^6\frac{-1}{2m_W^2}C^{W\nu_\delta\bar{l}_i}_LC^{W^*l_j\bar{\nu}_\delta}_LF_5(x_{\nu_\delta},x_W),
\nonumber\\&&A^L_2(W)=\sum_{\delta=1}^6\frac{1}{m_W^2}C^{W\nu_\delta\bar{l}_i}_LC^{W^*l_j\bar{\nu}_\delta}_L(1+\frac{m_{l_i}}{m_{l_j}})
F_6(x_{\nu_\delta},x_W),
\nonumber\\&&A_\alpha^{R}(W)=0,~~~\alpha=1,2.\label{oneloopW}
\end{eqnarray}
with
\begin{eqnarray}
&&F_5(x,y)=\frac{1}{96\pi^2}\Big[\frac{19xy-20x^2-5y^2}{(x-y)^3}+\frac{6x^2(2x-y)(\ln{x}-\ln{y})}{(x-y)^4}\Big],\nonumber\\
&&F_6(x,y)=\frac{1}{288\pi^2}\Big[\frac{65xy-43x^2-16y^2}{(x-y)^3}+\frac{6x^2(5x-6y)(\ln{x}-\ln{y})}{(x-y)^4}\Big].
\end{eqnarray}

The final Wilson coefficients are obtained by summing the expressions in Eqs.(\ref{oneloopn})(\ref{oneloopc})(\ref{oneloopW}) and the decay width for $l_j\rightarrow l_i\gamma$ can be expressed by Eq.(\ref{amplitude-gamma}):
\begin{eqnarray}
&&A^{L,R}_{\alpha}=A^{L,R}_{\alpha}(n)+A^{L,R}_{\alpha}(c)+A_\alpha^{L,R}(W),~~~\alpha=1,2,\nonumber\\
&&\Gamma(l_j\rightarrow{l_i\gamma})=\frac{e^2}{16\pi}m^5_{l_j}(|A^L_2|^2+|A^R_2|^2).
\end{eqnarray}

Finally, we get the branching ratio of $l_j\rightarrow{l_i\gamma}$:
\begin{eqnarray}
Br(l_j\rightarrow{l_i\gamma})=\frac{\Gamma(l_j\rightarrow{l_i\gamma})}{\Gamma_{l_j}}.
\end{eqnarray}

\subsection{$l_j \rightarrow 3l_i$}

\begin{figure}[h]
\setlength{\unitlength}{5mm}
\centering
\includegraphics[width=5.0in]{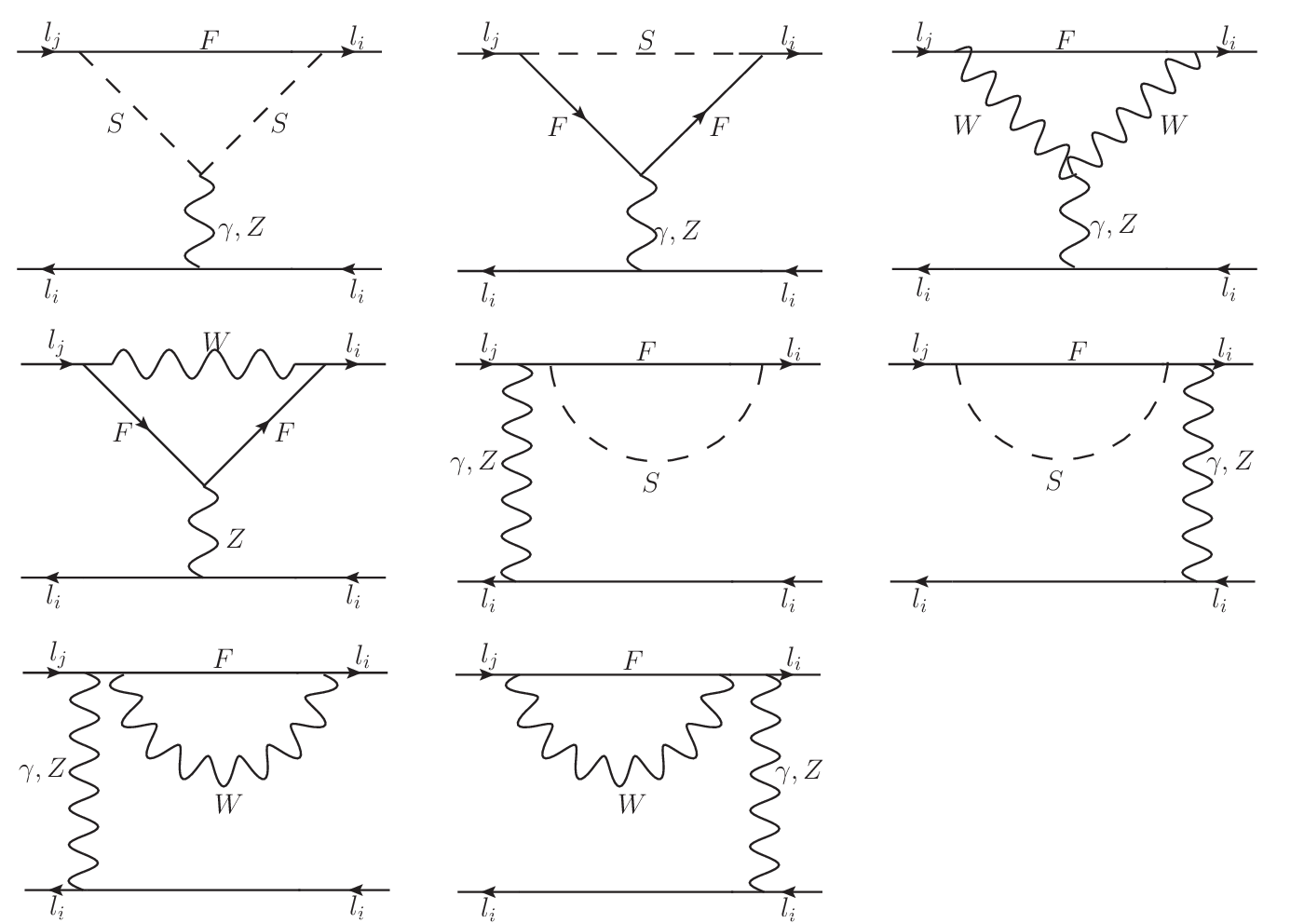}
\caption{ The penguin-type and self-energy diagrams for $l_j \rightarrow 3l_i$, with $F$ representing Dirac (Majorana) particles.}\label{N2}
\end{figure}

The effective Lagrangian for $l_j\rightarrow 3l_i$ decay processes receive contributions from penguin-type ($\gamma$-penguin and Z-penguin), self-energy diagrams and box-type diagrams. Let's first analyze the impact of the penguin-type and self-energy diagrams illustrated in Fig.\ref{N2}. on this transition mechanism.

Using Eq.(\ref{amplitude-gamma}), the contributions from the $\gamma$-penguin can be derived and expressed in the following form:
\begin{eqnarray}
&&T_{\gamma-p}=\overline{u}_i(p_1)[q^2\gamma_\mu(A^L_1P_L+A^R_1P_R)+m_{l_j}i\sigma_{\mu\nu}q^\nu(A^L_2P_L+A^R_2P_R)]u_j(p)\nonumber\\
&&~~~~~~~~\times\frac{e^2}{q^2}\overline{u}_i(p_2)\gamma^{\mu}v_i(p_3)-(p_1\leftrightarrow{p_2}).
\end{eqnarray}

The contributions from $Z$-penguin diagrams are derived following the same procedure as that used for the $\gamma$-penguin diagrams.
\begin{eqnarray}
&&T_{Z-p}=\frac{e^2}{m^2_Z}\overline{u}_i(p_1)\gamma_\mu(D_LP_L+D_RP_R)u_j(p)\overline{u}_i(p_2)\gamma^\mu(C^{Zl_i\overline{l}_i}_LP_L+C^{Zl_i\bar{l}_i}_RP_R)v_i(p_3)\nonumber\\
&&~~~~~~~~~-(p_1\leftrightarrow{p_2}).\nonumber\\
&&{D_{L,R}} = D_{L,R}(S) +D_{L,R}(W)\:.
\end{eqnarray}

The explicit expressions for the effective couplings are given by

\begin{eqnarray}
&&D_L(S)=\frac{1}{2{e^2}}\Big\{\Big[\sum_{\alpha=1}^6\sum_{\beta,n=1}^8\Big(\frac{2{m_{\chi^0_\beta}}{m_{\chi^0_n}}}
{{m_W^2}}C_R^{\tilde{L}_\alpha\chi^0_n\bar{l}_i}C_L^{Z\chi^0_\beta\bar{\chi}^0_n}
C_L^{\tilde{L}^*_\alpha{l_j}\bar{\chi}^0_\beta}{E_1}({x_{\tilde{L}_\alpha}},{x_{\chi^0_n}},{x_{\chi^0_\beta}})\nonumber\\&&\hspace{1.8cm}
-C_R^{\tilde{L}_\alpha\chi^0_n\bar{l}_i}C_R^{Z\chi^0_\beta\bar{\chi}^0_n}
C_L^{\tilde{L}^*_\alpha{l_j}\bar{\chi}^0_\beta}{E_2}({x_{\tilde{L}_\alpha}},{x_{\chi^0_n}},{x_{\chi^0_\beta}})\Big)\nonumber\\&&\hspace{1.8cm}
+\sum_{\kappa=1}^8\sum_{\rho,\sigma=1}^6C_R^{\tilde{L}_\sigma\chi^0_\kappa\bar{l}_i}C^{Z\tilde{L}_\rho\tilde{L}^*_\sigma}C_L^{{\tilde{L}^*_\rho}{l _j}{\bar{\chi}^0_\kappa}}{E_2}
({x_{\chi^0_\kappa}},{x_{\tilde{L}_\rho}},{x_{\tilde{L}_\sigma}})\Big]\nonumber\\&&\hspace{1.8cm}
+\Big[\sum_{\zeta=1}^6\sum_{\eta,\theta=1}^2\Big(\frac{2{m_{\chi^{\pm}_\eta}}{m_{\chi^{\pm}_\theta}}}
{{m_W^2}}C_R^{\tilde{\nu}_\zeta\chi^{\pm}_\theta\bar{l}_i}C_L^{Z\chi^{\pm}_\eta\bar{\chi}^{\pm}_\theta}
C_L^{\tilde{\nu}^*_\zeta{l_j}\bar{\chi}^{\pm}_\eta}{E_1}({x_{\tilde{\nu}_\zeta}},{x_{\chi^{\pm}_\theta}},{x_{\chi^{\pm}_\eta}})\nonumber\\&&\hspace{1.8cm}
-C_R^{\tilde{\nu}_\zeta\chi^{\pm}_\theta\bar{l}_i}C_R^{Z\chi^{\pm}_\eta\bar{\chi}^{\pm}_\theta}
C_L^{\tilde{\nu}^*_\zeta{l_j}\bar{\chi}^{\pm}_\eta}{E_2}({x_{\tilde{\nu}_\zeta}},{x_{\chi^{\pm}_\theta}},{x_{\chi^{\pm}_\eta}})\Big)\nonumber\\&&\hspace{1.8cm}
+\sum_{\xi=1}^2\sum_{\iota,\epsilon=1}^6C_R^{\tilde{\nu}_\epsilon\chi^{\pm}_\xi\bar{l}_i}C^{Z\tilde{\nu}_\iota\tilde{\nu}^*_\epsilon}C_L^{{\tilde{\nu}^*_\iota}{l _j}{\bar{\chi}^{\pm}_\xi}}{E_2}
({x_{\chi^{\pm}_\xi}},{x_{\tilde{\nu}_\iota}},{x_{\tilde{\nu}_\epsilon}})\Big]\nonumber\\&&\hspace{1.8cm}
+\Big[\sum_{\vartheta=1}^2\sum_{\delta,\varpi=1}^6\Big(\frac{2{m_{\nu_\delta}}{m_{\nu_\varpi}}}
{{m_W^2}}C_R^{H^{\pm}_\vartheta\nu_\varpi\bar{l}_i}C_L^{Z\nu_\delta\bar{\nu}_\varpi}
C_L^{H^{\pm*}_\vartheta{l_j}\bar{\nu}_\delta}{E_1}({x_{H^\pm_\vartheta}},{x_{\nu_\varpi}},{x_{\nu_\delta}})\nonumber\\&&\hspace{1.8cm}
-C_R^{H^{\pm}_\vartheta\nu_\varpi\bar{l}_i}C_R^{Z\nu_\delta\bar{\nu}_\varpi}
C_L^{H^{\pm*}_\vartheta{l_j}\bar{\nu}_\delta}{E_2}({x_{H^\pm_\vartheta}},{x_{\nu_\varpi}},{x_{\nu_\delta}})\Big)\nonumber\\&&\hspace{1.8cm}
+\sum_{\varrho=1}^6\sum_{\phi,\varphi=1}^2C_R^{H^\pm_\varphi\nu_\varrho\bar{l}_i}C^{ZH^\pm_\phi H^{\pm*}_\varphi}C_L^{{H^{\pm*}_\phi}{l _j}{\bar{\nu}_\varrho}}{E_2}
({x_{\nu_\varrho}},{x_{H^\pm_\phi}},{x_{H^\pm_\varphi}})\Big]\Big\},\nonumber\\
&&D_R(S) = \left. {D_L}(S) \right|{ _{L \leftrightarrow R}} \:.
\end{eqnarray}

The functions $E_1(x,y,z)$ and $E_2(x,y,z)$ are defined as follows:
\begin{eqnarray}
&&E_1(x,y,z)=\frac{1}{16\pi^2}\Big[\frac{x\ln{x}}{(x-y)(x-z)}+\frac{y\ln{y}}{(y-x)(y-z)}+\frac{z\ln{z}}{(z-x)(z-y)}\Big],\nonumber\\
&&E_2(x,y,z)=\frac{1}{16\pi^2}\Big[-(\Delta+1+\ln{x_\mu})+\frac{x^2\ln{x}}{(x-y)(x-z)}\nonumber\\
&&~~~~~~~~~~~~~~~~~~~~~~~~~~+\frac{y^2\ln{y}}{(y-x)(y-z)}+\frac{z^2\ln{z}}{(z-x)(z-y)}\Big].
\end{eqnarray}

We keep the small $m_i$ term and calculate the specific form of $D_{L,R}(W)$,
\begin{eqnarray}
&&D_L(W)=\frac{c_W}{es_W}\sum_{\delta=1}^6C^{W\nu_\delta\bar{l}_i}_LC^{W^*l_j\bar{\nu}_\delta}_L
\Big[E_3(x_{\nu_\delta},x_W)+2(x_i+x_j)E_6(x_{\nu_\delta},x_W)\Big]\nonumber\\&&\hspace{1.8cm}
+\frac{1}{e^2}\sum_{\varpi,\varrho=1}^6C^{W\nu_\varrho\bar{l}_i}_LC^{W^*l_j\bar{\nu}_\varpi}_LC^{Z^*\nu_\varpi\bar{\nu}_\varrho}_L
\Big\{-\frac{3}{32\pi^2}-E_2(x_W,x_{\nu_\varpi},x_{\nu_\varrho})\nonumber\\&&\hspace{1.8cm}+x_j\Big[\frac{1}{3}E_4(x_W,x_{\nu_\varpi},x_{\nu_\varrho})
+E_5(x_W,x_{\nu_\varpi},x_{\nu_\varrho})\Big]\Big\},\nonumber\\&&
D_R(W)=\frac{c_W}{es_W}\sum_{\delta=1}^6C^{W\nu_\delta\bar{l}_i}_LC^{W^*l_j\bar{\nu}_\delta}_L2\sqrt{x_ix_j}E_6(x_{\nu_\delta},x_W)
\nonumber\\&&\hspace{1.8cm}+\frac{1}{e^2}\sum_{\varpi,\varrho=1}^6C^{W\nu_\varrho\bar{l}_i}_Lc^{W^*l_j\bar{\nu}_\varpi}_LC^{Z^*\nu_\varpi\bar{\nu}_\varrho}_L\sqrt{x_ix_j}
\Big[2E_1(x_W,x_{\nu_\varpi},x_{\nu_\varrho})\nonumber\\&&\hspace{1.8cm}
-\frac{1}{3}E_4(x_W,x_{\nu_\varpi},x_{\nu_\varrho})
-2E_5(x_W,x_{\nu_\varpi},x_{\nu_\varrho})  \Big].
\end{eqnarray}

Here, $x_i= {m_{l_i}^2}/{m_W^2}$ and $x_j= {m_{l_j}^2}/{m_W^2}$.

The concrete expressions for the functions $E_3(x,y),E_4(x,y,z), E_5(x,y,z)$ and $E_6(x,y)$ are collected here
\begin{eqnarray}
&&E_3(x,y)=\frac{-1}{16 \pi ^2}\Big[(\Delta+\ln x_\mu+1)+\frac{y^2\ln y-x^2 \ln x}{(y-x)^2}+\frac{y+2 y\ln y}{x-y}-\frac{1}{2}\Big],\nonumber\\&&
E_4(x,y,z)=\frac{1}{32\pi^2}\Big[\frac{2x^3[3x(x-y-z)+y^2+yz+z^2]\ln x}{(x-y)^3(x-z)^3}\nonumber\\&&\hspace{2.6cm}-\frac{2(3x^2-3xy+y^2)y\ln y}{(x-y)^3(y-z)}+\frac{2(3x^2-3xz+z^2)z\ln z}{(x-z)^3(y-z)}\nonumber\\&&\hspace{2.6cm}
-\frac{x[5x^2-3x(y+z)+yz]}{(x-y)^2(x-z)^2}
\Big],\nonumber\\
&&E_5(x,y,z)=\frac{1}{16\pi^2}\Big[\frac{x^2(2x-y-z)\ln x}{(x-y)^2(x-z)^2}+\frac{y(y-2x)\ln y}{(x-y)^2(y-z)}
\nonumber\\&&\hspace{2.6cm}-\frac{x}{(x-y)(x-z)}+\frac{z(2x-z)\ln z}{(x-z)^2(y-z)}\Big],\nonumber\\&&
E_6(x,y)=\frac{1}{96\pi^2}\Big[\frac{6x^2y(\ln{x}-\ln{y})}{(x-y)^4}+\frac{-5xy-2x^2+y^2}{(x-y)^3}\Big].
\end{eqnarray}

\begin{figure}[h]
\setlength{\unitlength}{5.0mm}
\centering
\includegraphics[width=5.0in]{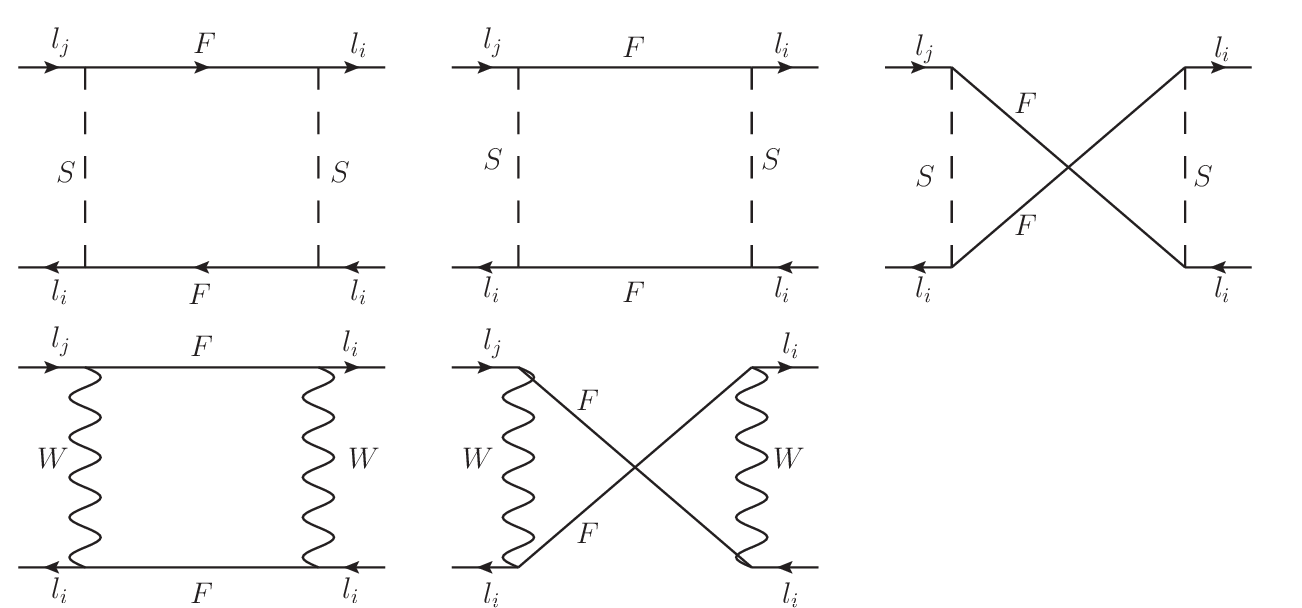}
\caption{ The box-type diagrams for $l_j\rightarrow 3l_i$, with $F$ representing Dirac (Majorana) particles.}\label{N3}
\end{figure}
The box-type diagrams depicted in Fig.\ref{N3} can be expressed in the following form:
\begin{eqnarray}
&&T_{box} = \Big\{B_1^L{e^2}{\bar u_i}({p_1}){\gamma _\mu }{P_L}{u_j}(p){\bar u_i}(p_2)
{\gamma ^\mu }{P_L}{v_i}({p_3}) + (L \leftrightarrow R)\Big\}   \nonumber\\
&&\quad + \: \Big\{B_2^L{e^2}\Big[{\bar u_i}(p_1){\gamma _\mu }{P_L}{u_j}(p){\bar u_i}(p_2)
{\gamma ^\mu }{P_R}{v_i}({p_3}) - (p_1 \leftrightarrow {p_2})\Big] + {(L \leftrightarrow R) } \Big\} \nonumber\\
&&\quad + \: \Big\{ B_3^L{e^2}\Big[{\bar u_i}({p_1}){P_L}{u_j}(p){\bar u_i}({p_2})
{P_L}{v_i}({p_3}) - ({p_1} \leftrightarrow {p_2})\Big] + (L \leftrightarrow R) \Big\}   \nonumber\\
&&\quad + \: \Big\{ B_4^L{e^2}\Big[{\bar u_i}({p_1}){\sigma _{\mu \nu }}{P_L}{u_j}(p){\bar u_i}({p_2}){\sigma ^{\mu \nu }}{P_L}{v_i}({p_3})
- ({p_1} \leftrightarrow {p_2})\Big]  + \: (L \leftrightarrow R)\Big\} \:
\nonumber\\&&\quad + \: \Big\{ B_5^L{e^2}\Big[{\bar u_i}({p_1}){P_L}{u_j}(p){\bar u_i}({p_2}){P_R}{v_i}({p_3})
- ({p_1} \leftrightarrow {p_2})\Big]  + \: (L \leftrightarrow R)\Big\}.
\end{eqnarray}

The virtual chargino contributions to the effective couplings $B^{L,R}_\beta(c),~\beta=1...5$ are determined from the box-type diagrams.
\begin{eqnarray}
&&B^L_1(c)=\sum_{\eta,\theta=1}^2\sum_{\epsilon,\iota=1}^6\frac{1}{2e^2m^2_W}C^{{\tilde{\nu}_\iota}{\chi^\pm_\eta}\bar{l}_i}_RC^{{\tilde{\nu}_\epsilon}l_j\bar{\chi}^\pm_\eta}_LC^{{\tilde{\nu}_\epsilon}{\chi^\pm_\theta}\bar{l}_i}_RC^{{\tilde{\nu}_\iota}l_i\bar{\chi}^\pm_\theta}_LE_7(x_{\chi^\pm_\eta},x_{\chi^\pm_\theta},x_{\tilde{\nu}_\epsilon},x_{\tilde{\nu}_\iota}),\nonumber\\
&&B^L_2(c)=\sum_{\eta,\theta=1}^2\sum_{\epsilon,\iota=1}^6\Big[\frac{1}{4e^2m^2_W}C^{{\tilde{\nu}_\iota}{\chi^\pm_\eta}\bar{l}_i}_RC^{{\tilde{\nu}_\epsilon}l_j\bar{\chi}^\pm_\eta}_LC^{{\tilde{\nu}_\epsilon}{\chi^\pm_\theta}\bar{l}_i}_LC^{{\tilde{\nu}_\iota}l_i\bar{\chi}^\pm_\theta}_RE_7(x_{\chi^\pm_\eta},x_{\chi^\pm_\theta},x_{\tilde{\nu}_\epsilon},x_{\tilde{\nu}_\iota}),\nonumber\\
&&~~~~~~~~~-\frac{m_{\chi^\pm_\eta}m_{\chi^\pm_\theta}}{2e^2m^4_W}C^{{\tilde{\nu}_\iota}{\chi^\pm_\eta}\bar{l}_i}_RC^{{\tilde{\nu}_\epsilon}l_j\bar{\chi}^\pm_\eta}_RC^{{\tilde{\nu}_\epsilon}{\chi^\pm_\theta}\bar{l}_i}_LC^{{\tilde{\nu}_\iota}l_i\bar{\chi}^\pm_\theta}_LE_8(x_{\chi^\pm_\eta},x_{\chi^\pm_\theta},x_{\tilde{\nu}_\epsilon},x_{\tilde{\nu}_\iota})\Big],\nonumber\\
&&B^L_3(c)=\sum_{\eta,\theta=1}^2\sum_{\epsilon,\iota=1}^6\frac{m_{\chi^\pm_\eta}m_{\chi^\pm_\theta}}{e^2m^4_W}C^{{\tilde{\nu}_\iota}{\chi^\pm_\eta}\bar{l}_i}_LC^{{\tilde{\nu}_\epsilon}l_j\bar{\chi}^\pm_\eta}_LC^{{\tilde{\nu}_\epsilon}{\chi^\pm_\theta}\bar{l}_i}_LC^{{\tilde{\nu}_\iota}l_i\bar{\chi}^\pm_\theta}_LE_8(x_{\chi^\pm_\eta},x_{\chi^\pm_\theta},x_{\tilde{\nu}_\epsilon},x_{\tilde{\nu}_\iota}),\nonumber\\
&&B^L_4(c)=B^L_5(c)=0,\nonumber\\
&&B^R_\beta(c)=B^L_\beta(c)|_{L\leftrightarrow{R}},~~~\beta=1...5,\label{B4C}
\end{eqnarray}
with
\begin{eqnarray}
&&{E_7}(x , y , z, t) = \frac{1}{{16{\pi ^2}}}\Big[\frac{{x^2\ln {x}}}{{({x} - {y})({x} - {z})({x} - {t})}}  +\; \frac{{y^2\ln {y}}}{{({y} - {x})({y} - {z})({y} - {t})}}\nonumber\\
&&\hspace{3.0cm}  + \frac{{z^2\ln {z}}}{{({z}  - {x})({z} - {y})({z} - {t})}} + \frac{{t^2\ln {t}}}{{({t} - {x})({t} - {y})({t} - {z})}}\Big]\:,
\nonumber\\&&
{E_8}(x , y , z, t) = \frac{1}{{16{\pi ^2}}}\Big[\frac{{{x}\ln {x}}}{{({x} - {y})({x} - {z})({x} - {t})}}  +\frac{{{y}\ln {y}}}{{({y} - {x})({y} - {z})({y} - {t})}} \nonumber\\
&&\hspace{3.0cm} + \frac{{{z}\ln {z}}}{{({z}  - {x})({z} - {y})({z} - {t})}}  + \: \frac{{{t}\ln {t}}}{{({t} - {x})({t} - {y})({t} - {z})}}\Big] .
\end{eqnarray}

For the box-type diagrams, the neutralino-slepton, neutrino-charged Higgs and lepton neutralino-slepton contributions to the effective couplings $B_\beta^{L,R}(n),~\beta=1...5$ are written,
\begin{eqnarray}
&&B^L_1(n)=\sum_{\beta,n=1}^8\sum_{\rho,\sigma=1}^6\Big\{\frac{m_{\chi^0_\beta}m_{\chi^0_n}}{e^2m^4_W}E_8(x_{\chi^0_\beta},x_{\chi^0_n},x_{\tilde{L}_\rho},x_{\tilde{L}_\sigma})C^{{\tilde{L}_\sigma}{\chi^0_\beta}\bar{l}_i}_LC^{{\tilde{L}}^*_{\rho}l_j\bar{\chi}^0_\beta}_LC^{{\tilde{L}_\sigma}{\chi^0_n}\bar{l}_i}_RC^{{\tilde{L}}^*_{\rho}l_i\bar{\chi}^0_n}_R\nonumber\\
&&~~~~~~~~~~~+\frac{1}{2e^2m^2_W}E_7(x_{\chi^0_\beta},x_{\chi^0_n},x_{\tilde{L}_\rho},x_{\tilde{L}_\sigma})\Big[C^{{\tilde{L}_\sigma}{\chi^0_\beta}\bar{l}_i}_RC^{{\tilde{L}}^*_{\rho}l_j\bar{\chi}^0_\beta}_LC^{{\tilde{L}_\rho}{\chi^0_n}\bar{l}_i}_RC^{{\tilde{L}}^*_{\sigma}l_i\bar{\chi}^0_n}_L\nonumber\\
&&~~~~~~~~~~~+C^{{\tilde{L}_\sigma}{\chi^0_\beta}\bar{l}_i}_LC^{{\tilde{L}}^*_{\rho}l_j\bar{\chi}^0_\beta}_RC^{{\tilde{L}_\sigma}{\chi^0_n}\bar{l}_i}_RC^{{\tilde{L}}^*_{\rho}l_i\bar{\chi}^0_n}_L\Big]\Big\}\nonumber\\
&&~~~~~~~~~~~+\sum_{\varpi,\varrho=1}^6\sum_{\phi,\varphi=1}^2\Big\{\frac{m_{\nu_\varpi}m_{\nu_\varrho}}{e^2m^4_W}E_8(x_{\nu_\varpi},x_{\nu\varrho},x_{H^{\pm}_\phi},x_{H^{\pm}_\varphi})C^{{H^{\pm}_\varphi}{\nu_\varpi}\bar{l}_i}_LC^{H^{\pm*}_\phi l_j\bar{\nu}_\varpi}_LC^{{H^{\pm}_\varphi}{\nu_\varrho}\bar{l}_i}_RC^{H^{\pm*}_\phi l_i\bar{\nu}_\varrho}_R\nonumber\\
&&~~~~~~~~~~~+\frac{1}{2e^2m^2_W}E_7(x_{\nu_\varpi},x_{\nu\varrho},x_{H^{\pm}_\phi},x_{H^{\pm}_\varphi})\Big[C^{{H^{\pm}_\varphi}{\nu_\varpi}\bar{l}_i}_RC^{H^{\pm*}_\phi l_j\bar{\nu}_\varpi}_LC^{{H^{\pm}_\phi}{\nu_\varrho}\bar{l}_i}_RC^{H^{\pm*}_{\varphi}l_i\bar{\nu}_\varrho}_L\nonumber\\
&&~~~~~~~~~~~+C^{{H^{\pm}_\varphi}{\nu_\varpi}\bar{l}_i}_LC^{H^{\pm*}_\phi l_j\bar{\nu}_\varpi}_RC^{{H^{\pm}_\varphi}{\nu_\varrho}\bar{l}_i}_RC^{H^{\pm*}_\phi l_i\bar{\nu}_\varrho}_L\Big]\Big\},\nonumber\\
&&B^L_2(n)=\sum_{\beta,n=1}^8\sum_{\rho,\sigma=1}^6\Big\{-\frac{m_{\chi^0_\beta}m_{\chi^0_n}}{2e^2m^4_W}E_8(x_{\chi^0_\beta},x_{\chi^0_n},x_{\tilde{L}_\rho},x_{\tilde{L}_\sigma})C^{{\tilde{L}_\sigma}{\chi^0_\beta}\bar{l}_i}_RC^{{\tilde{L}}^*_{\rho}l_j\bar{\chi}^0_\beta}_RC^{{\tilde{L}_\rho}{\chi^0_n}\bar{l}_i}_LC^{{\tilde{L}}^*_{\sigma}l_i\bar{\chi}^0_n}_L\nonumber\\
&&~~~~~~~~~~~+\frac{1}{4e^2m^2_W}E_7(x_{\chi^0_\beta},x_{\chi^0_n},x_{\tilde{L}_\rho},x_{\tilde{L}_\sigma})\Big[C^{{\tilde{L}_\sigma}{\chi^0_\beta}\bar{l}_i}_RC^{{\tilde{L}}^*_{\rho}l_j\bar{\chi}^0_\beta}_LC^{{\tilde{L}_\rho}{\chi^0_n}\bar{l}_i}_LC^{{\tilde{L}}^*_{\sigma}l_i\bar{\chi}^0_n}_R\nonumber\\
&&~~~~~~~~~~~+C^{{\tilde{L}_\sigma}{\chi^0_\beta}\bar{l}_i}_RC^{{\tilde{L}}^*_{\rho}l_j\bar{\chi}^0_\beta}_LC^{{\tilde{L}_\sigma}{\chi^0_n}\bar{l}_i}_RC^{{\tilde{L}}^*_{\rho}l_i\bar{\chi}^0_n}_L\Big]\Big\}\nonumber\\
&&~~~~~~~~~~~+\sum_{\varpi,\varrho=1}^6\sum_{\phi,\varphi=1}^2\Big\{-\frac{m_{\nu_\varpi}m_{\nu_\varrho}}{2e^2m^4_W}E_8(x_{\nu_\varpi},x_{\nu\varrho},x_{H^{\pm}_\phi},x_{H^{\pm}_\varphi})C^{{H^{\pm}_\varphi}{\nu_\varpi}\bar{l}_i}_RC^{H^{\pm*}_\phi l_j\bar{\nu}_\varpi}_RC^{{H^{\pm}_\phi}{\nu_\varrho}\bar{l}_i}_LC^{H^{\pm*}_\varphi l_i\bar{\nu}_\varrho}_L\nonumber\\
&&~~~~~~~~~~~+\frac{1}{4e^2m^2_W}E_7(x_{\nu_\varpi},x_{\nu\varrho},x_{H^{\pm}_\phi},x_{H^{\pm}_\varphi})\Big[C^{{H^{\pm}_\varphi}{\nu_\varpi}\bar{l}_i}_RC^{H^{\pm*}_\phi l_j\bar{\nu}_\varpi}_LC^{{H^{\pm}_\phi}{\nu_\varrho}\bar{l}_i}_LC^{H^{\pm*}_{\varphi}l_i\bar{\nu}_\varrho}_R\nonumber\\
&&~~~~~~~~~~~+C^{{H^{\pm}_\varphi}{\nu_\varpi}\bar{l}_i}_RC^{H^{\pm*}_\phi l_j\bar{\nu}_\varpi}_LC^{{H^{\pm}_\varphi}{\nu_\varrho}\bar{l}_i}_RC^{H^{\pm*}_\phi l_i\bar{\nu}_\varrho}_L\Big]\Big\},\nonumber\\
&&B^L_3(n)=\sum_{\beta,n=1}^8\sum_{\rho,\sigma=1}^6\frac{m_{\chi^0_\beta}m_{\chi^0_n}}{e^2m^4_W}E_8(x_{\chi^0_\beta},x_{\chi^0_n},x_{\tilde{L}_\rho},x_{\tilde{L}_\sigma})\Big[C^{{\tilde{L}_\sigma}{\chi^0_\beta}\bar{l}_i}_LC^{{\tilde{L}}^*_{\rho}l_j\bar{\chi}^0_\beta}_LC^{{\tilde{L}_\rho}{\chi^0_n}\bar{l}_i}_LC^{{\tilde{L}}^*_{\sigma}l_i\bar{\chi}^0_n}_L\nonumber\\
&&~~~~~~~~~~~-\frac{1}{2}C^{{\tilde{L}_\sigma}{\chi^0_\beta}\bar{l}_i}_LC^{{\tilde{L}}^*_{\rho}l_j\bar{\chi}^0_\beta}_LC^{{\tilde{L}_\sigma}{\chi^0_n}\bar{l}_i}_LC^{{\tilde{L}}^*_{\rho}l_i\bar{\chi}^0_n}_L\Big]\nonumber\\
&&~~~~~~~~~~~+\sum_{\varpi,\varrho=1}^6\sum_{\phi,\varphi=1}^2\frac{m_{\nu_\varpi}m_{\nu_\varrho}}{e^2m^4_W}E_8(x_{\nu_\varpi},x_{\nu\varrho},x_{H^{\pm}_\phi},x_{H^{\pm}_\varphi})\Big[C^{{H^{\pm}_\varphi}{\nu_\varpi}\bar{l}_i}_LC^{H^{\pm*}_\phi l_j\bar{\nu}_\varpi}_LC^{{H^{\pm}_\phi}{\nu_\varrho}\bar{l}_i}_LC^{H^{\pm*}_{\varphi}l_i\bar{\nu}_\varrho}_L\nonumber\\
&&~~~~~~~~~~~-\frac{1}{2}C^{{H^{\pm}_\varphi}{\nu_\varpi}\bar{l}_i}_LC^{H^{\pm*}_\phi l_j\bar{\nu}_\varpi}_LC^{{H^{\pm}_\varphi}{\nu_\varrho}\bar{l}_i}_LC^{H^{\pm*}_\phi l_i\bar{\nu}_\varrho}_L\Big],\nonumber\\
&&B^L_4(n)=\sum_{\beta,n=1}^8\sum_{\rho,\sigma=1}^6\frac{m_{\chi^0_\beta}m_{\chi^0_n}}{8e^2m^4_W}E_8(x_{\chi^0_\beta},x_{\chi^0_n},x_{\tilde{L}_\rho},x_{\tilde{L}_\sigma})C^{{\tilde{L}_\sigma}{\chi^0_\beta}\bar{l}_i}_LC^{{\tilde{L}}^*_{\rho}l_j\bar{\chi}^0_\beta}_LC^{{\tilde{L}_\sigma}{\chi^0_n}\bar{l}_i}_LC^{{\tilde{L}}^*_{\rho}l_i\bar{\chi}^0_n}_L\nonumber\\
&&~~~~~~~~~~~+\sum_{\varpi,\varrho=1}^6\sum_{\phi,\varphi=1}^2\frac{m_{\nu_\varpi}m_{\nu_\varrho}}{8e^2m^4_W}E_8(x_{\nu_\varpi},x_{\nu\varrho},x_{H^{\pm}_\phi},x_{H^{\pm}_\varphi})C^{{H^{\pm}_\varphi}{\nu_\varpi}\bar{l}_i}_LC^{H^{\pm*}_\phi l_j\bar{\nu}_\varpi}_LC^{{H^{\pm}_\varphi}{\nu_\varrho}\bar{l}_i}_LC^{H^{\pm*}_\phi l_i\bar{\nu}_\varrho}_L,\nonumber\\
&&B^L_5(n)=0,\nonumber\\
&&B^R_\beta(n)=B^L_\beta(n)|_{L\leftrightarrow{R}},~~~\beta=1...5,\label{B4N}
\end{eqnarray}

We also deduce the box-type contributions from virtual W-neutrino
\begin{eqnarray}
&&B_1^{L}(W) = \sum_{\varpi,\varrho=1}^6\frac{1}{{e^2}{m_W^2}}\Big[-\frac{\partial }{\partial x_W}
{E_2}(x_{W},x_{\nu_\varpi},x_{\nu_\varrho})C_L^{Wl_j\bar{\nu}_\varpi}C_L^{W^*\nu_\varpi\bar{l}_i}C_L^{W^*\nu_\varrho\bar{l}_i}C_L^{Wl_i\bar{\nu}_\varrho}\nonumber\\&&
~~~~~~~~~~-2\frac{m_{\nu_\varpi}m_{\nu_\varrho}}{{m_W^2}}\frac{\partial }{\partial x_W}
{E_1}(x_{W},x_{\nu_\varpi},x_{\nu_\varrho})C_L^{Wl_j\bar{\nu}_\varpi}C_L^{W^*\nu_\varrho\bar{l}_i}C_L^{Wl_i\bar{\nu}_\varrho}C_L^{W^*\nu_\varpi\bar{l}_i}\Big],
\nonumber\\&&
B_3^L(W)=\sum_{\varpi,\varrho=1}^6-\frac{7}{2}\frac{m_{\nu_\varpi}m_{\nu_\varrho}}{{e^2m_W^4}}\frac{\partial }{\partial x_W}
{E_1}(x_{W},x_{\nu_\varpi},x_{\nu_\varrho})C_L^{Wl_j\bar{\nu}_\varpi}C_L^{W^*\nu_\varrho\bar{l}_i}C_L^{Wl_i\bar{\nu}_\varrho}C_L^{W^*\nu_\varpi\bar{l}_i},\nonumber\\&&
B_2^L(W)=0,~~~B_4^L(W)=\frac{1}{14}B_3^L(W),~~B_5^L(W)=-\frac{1}{7}B_3^L(W),\nonumber\\&&
B_1^{R}(W)=0,~~~
B_\alpha^{R}(W) = {B_\alpha^{L}}(W)|_{L\leftrightarrow{R}},~~~\alpha=2...5,\label{B4W}
\end{eqnarray}

With Eqs.(\ref{B4C})(\ref{B4N})(\ref{B4W}), we formulate the coefficients $B_\beta^{L,R}$ as follows:
\begin{eqnarray}
B_\beta^{L,R} = B_\beta^{L,R}(c) + B_\beta^{L,R}(n)+ B_\beta^{L,R}(W),~~~(\beta = 1\dots5)\:.
\end{eqnarray}

The decay widths for $l_j  \rightarrow 3l_i$ can be calculated by evaluating the corresponding amplitudes,
\begin{eqnarray}
&&\Gamma (l_j  \to 3l_i) = \frac{{{e^4}}}{{512{\pi ^3}}}m_{{l_j}}^5  \Big\{
({\left| {A_2^L} \right|^2} + {\left| {A_2^R} \right|^2})(\frac{{16}}{3}\ln
\frac{{{m_{{l_j}}}}}{{2{m_{{l_i}}}}} - \frac{{14}}{9}) \nonumber\\
&&\quad + \: ({\left| {A_1^L} \right|^2} + {\left| {A_1^R} \right|^2}) - 2(A_1^LA_2^{R * }
 + A_2^LA_1^{R * } + A_2^LA_1^{R * }+A_1^LA_2^{R * }) + \frac{1}{6}({\left| {B_1^L} \right|^2}  + {\left| {B_1^R} \right|^2}) \nonumber\\
&&\quad  + \: \frac{1}{3}({\left| {B_2^L} \right|^2} + {\left| {B_2^R} \right|^2})
+ \frac{1}{{24}}({\left| {B_3^L} \right|^2} + {\left| {B_3^R} \right|^2}) + 6({\left| {B_4^L} \right|^2} + {\left| {B_4^R} \right|^2})+\frac{1}{12}({\left| {B_5^L} \right|^2} + {\left| {B_5^R} \right|^2})
 \nonumber\\
&&\quad- \frac{1}{6}(B_2^LB_5^{L * } + B_2^RB_5^{R * }
+A_1^LB_5^{L * } + A_1^RB_5^{R * }+ B_5^RB_2^{R * } + B_5^LB_2^{L * }
+B_5^RA_1^{R * } + B_5^LA_1^{L * }) \nonumber\\&&\quad
+\;\frac{1}{3}(A_2^RB_5^{L * } + A_2^LB_5^{R * }+ B_5^RA_2^{L * } + B_5^LA_2^{R * })- \frac{1}{6}(D_{LR}B_5^{L * } + D_{RL}B_5^{R * }
+ B_5^RD_{RL}^* + B_5^LD_{LR}^*)
\nonumber\\
&&\quad  - \: \frac{1}{2}(B_3^LB_4^{L * } + B_3^RB_4^{R * } + B_4^RB_3^{R * } + B_4^LB_3^{L * }) +
\frac{1}{3}(A_1^LB_1^{L * } + A_1^RB_1^{R * } + A_1^LB_2^{L * }  \nonumber\\
&&\quad  + \: A_1^RB_2^{R * } + B_1^RA_1^{R * } + B_1^LA_1^{L * } + B_2^RA_1^{R * }+ B_2^LA_1^{L * }) - \frac{2}{3}(A_2^RB_1^{L * } + A_2^LB_1^{R * }  +A_2^LB_2^{R * }\nonumber\\
&&\quad  + \: A_2^RB_2^{L * } + B_1^RA_2^{L * } + B_1^LA_2^{R * }  +B_2^LA_2^{R * }+B_2^RA_2^{L * })+\frac{1}{3}\Big[2({\left| {{D_{LL}}} \right|^2} + {\left| {{D_{RR}}} \right|^2})  \nonumber\\
&&\quad + \: ({\left| {{D_{LR}}} \right|^2} + {\left| {{D_{RL}}} \right|^2}) + (B_1^LD_{LL}^ *  + B_1^RD_{RR}^ *+B_2^LD_{LR}^ *  + B_2^RD_{RL}^ * + D_{RR}B_1^{R *}  \nonumber\\
&&\quad  + \:  D_{LL}B_1^{L *}+D_{RL}B_2^{R *}+D_{LR}B_2^{L *} ) + 2(A_1^LD_{LL}^ *  + A_1^RD_{RR}^ * +D_{RR}A_1^{R *}+D_{LL}A_1^{L *})\nonumber\\
&&\quad +\:  (A_1^LD_{LR}^ *  + A_1^RD_{RL}^ *  + D_{RL}A_1^{R *}+ D_{LR}A_1^{L *})-4(A_2^RD_{LL}^ *  + A_2^LD_{RR}^ *  + D_{RR}A_2^{L *} \nonumber\\
&&\quad+ \: D_{LL}A_2^{R *})- 2(A_2^LD_{RL}^ *  + A_2^RD_{LR}^ *  + D_{LR}A_2^{R *}+D_{RL}A_2^{L *})\Big]\Big\},
\end{eqnarray}
with
\begin{eqnarray}
&&D_{LL}=\frac{D_LC^{Zl_i\bar{l}_i}_L}{m^2_Z},~~~~~D_{RR}=D_{LL}|_{L\leftrightarrow{R}},\nonumber\\
&&D_{LR}=\frac{D_LC^{Zl_i\bar{l}_i}_R}{m^2_Z},~~~~~D_{RL}=D_{LR}|_{L\leftrightarrow{R}}.
\end{eqnarray}

Finally, the branching ratios of $l_j\rightarrow3l_i$ are given by
\begin{eqnarray}
Br(l_j\rightarrow3l_i)=\frac{\Gamma(l_j\rightarrow{3l_i})}{\Gamma_{l_j}}
\end{eqnarray}

\subsection{$\mu\rightarrow e+q\bar q$}

\begin{figure}
\setlength{\unitlength}{4.0mm}
\centering
\includegraphics[width=4.0in]{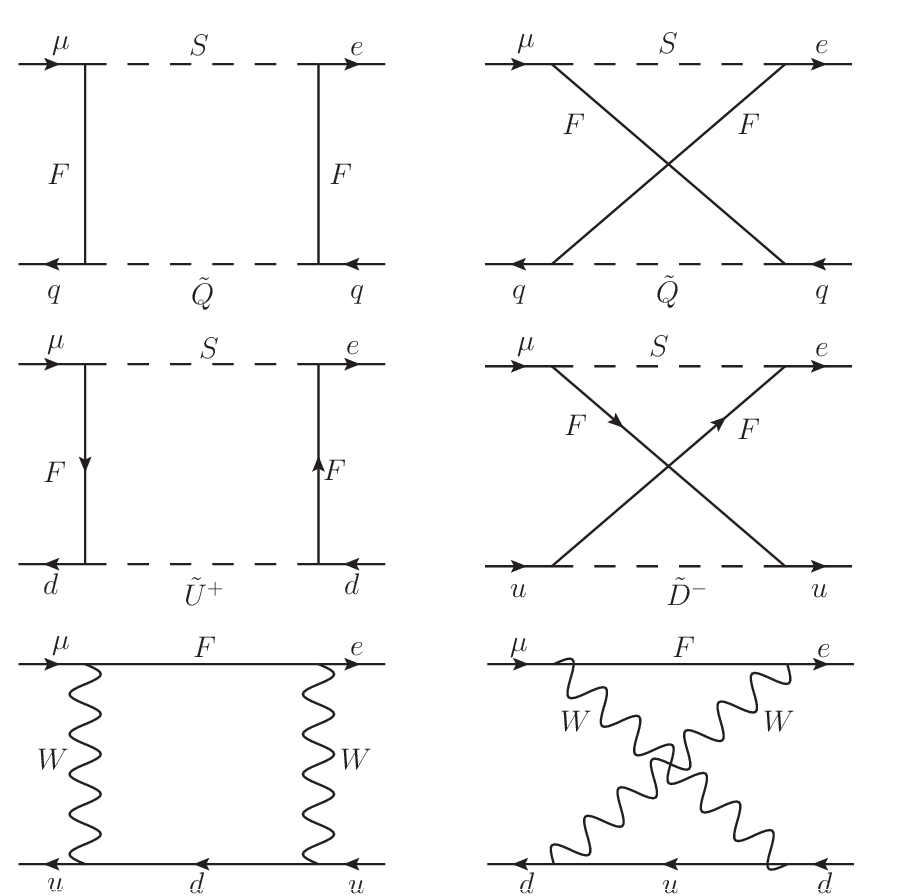}
\caption[]{The box-type diagrams for $\mu\rightarrow \emph{e}+\rm{q}\bar q$, with $F$ representing Dirac(Majorana) particles.}\label{FMFLV2}
\end{figure}

The effective Lagrangian corresponding to the box-type diagrams shown in Fig.\ref{FMFLV2} can be expressed as

\begin{eqnarray}
&& T_{{\rm{box}}}  = e^2 \sum_{q=u,d}\bar q\gamma _\alpha  q\bar e\gamma ^\alpha  \left( {O_q^L P_L  + O_q^R P_R } \right)\mu \:.
\end{eqnarray}
with
\begin{eqnarray}
&& O_q^{L,R}  = O_q^{(n)L,R}  + O_q^{(c)L,R}  + O_q^{(W)L,R}  (q=u,d) \:.
\end{eqnarray}

$O_q(n)$ represent the contributions from the virtual neutral Fermion diagrams in the first line of Fig.\ref{FMFLV2}.
\begin{eqnarray}
 && O_q^L (n) = \sum_{\beta,n=1}^8\sum_{\alpha,l=1}^6\Big\{\frac{1}{8e^2 m_W^2 }E_7( x_{\chi _\beta^0 } ,x_{\chi _n^0 } ,x_{\tilde L_\alpha} ,x_{\tilde Q_l} )
\Big[C_L^{\tilde L_\alpha \mu \bar{\chi} _\beta^0 }  C_R^{\tilde L_\alpha\chi _n^0 \bar e} C_R^{\tilde Q_l\chi _\beta^0 \bar q}
 C_L^{\tilde Q_l q \bar{\chi} _n^0 } \nonumber \\
 &&\hspace{1.8cm}- C_L^{\tilde L_\alpha \mu \bar{\chi} _\beta^0  } C_R^{\tilde L_\alpha\chi _n^0 \bar e}
 C_R^{\tilde Q_l q \bar{\chi} _\beta^0 }C_L^{\tilde Q_l\chi _n^0 \bar q} \Big]\nonumber \\
 &&\hspace{1.8cm} - \frac{{m_{\chi _\beta^0 } m_{\chi _n^0 } }}{{4e^2 m_W^4}}E_8
 ( x_{\chi _\beta^0 } ,x_{\chi _n^0 } ,x_{\tilde L_\alpha} ,x_{\tilde Q_l})
 \Big[C_L^{\tilde L_\alpha \mu \bar{\chi}_\beta^0  } C_R^{\tilde L_\alpha\chi _n^0 \bar e} C_L^{\tilde Q_l\chi _\beta^0 \bar q}
  C_R^{\tilde Q_l q \bar{\chi} _n^0 }  \nonumber \\
 &&\hspace{1.8cm} - C_L^{\tilde L_\alpha \mu \bar{\chi} _\beta^0  } C_R^{\tilde L_\alpha\chi _n^0 \bar e}
  C_L^{\tilde Q_l q \bar{\chi} _\beta^0 } C_R^{\tilde Q_l\chi _n^0 \bar q} \Big]\Big\}, \nonumber \\
 &&O_q^R (n) = O_q^L (n)|_{L \leftrightarrow R} , \;\;\left( q = u,d \right).
 \end{eqnarray}

Correspondingly, the virtual charged Fermion diagrams give contributions denoted by $O_q(c)$.
\begin{eqnarray}
&& O_d^L (c) = \sum_{\eta,\theta=1}^2\sum_{\sigma,l=1}^6\Big[ \frac{1}{{8e^2 m_W^2 }}E_7 ({x_{\chi _\eta^ \pm  } ,x_{\chi _\theta^ \pm  } ,x_{\tilde \nu_\sigma } ,x_{\tilde{U}_l^ +  } } )C_L^{\tilde \nu_\sigma \mu  \bar{\chi} _\eta^ \pm  } C_R^{\tilde \nu_\sigma \chi _\theta^ \pm  \bar e} C_R^{\tilde{U}_l^ +  \chi _\eta^ \pm  \bar d}
 C_L^{\tilde{U}_l^ + d \bar{\chi} _\theta^ \pm } \nonumber \\
  &&\hspace{1.8cm}- \frac{{m_{\chi _\eta^ \pm  } m_{\chi _\theta^ \pm  } }}{{4e^2 m_W^4 }}E_8 ({x_{\chi _\eta^ \pm  } ,x_{\chi _\theta^ \pm  } ,x_{\tilde \nu_\sigma } ,x_{\tilde{U}_l^ +}})C_L^{\tilde \nu_\sigma \mu \bar{\chi} _\eta^ \pm  } C_R^{\tilde \nu_\sigma \chi _\theta^ \pm  \bar e}
  C_L^{\tilde{U}_l^ +  \chi _\eta^ \pm  \bar d} C_R^{\tilde{U}_l^ + d \bar{\chi} _\theta^ \pm   }  \Big],\nonumber \\
&& O_u^L (c) =   \sum_{\eta,\theta=1}^2\sum_{\sigma,l=1}^6\Big[ \frac{-1}{{8e^2 m_W^2 }}
E_7 ({x_{\chi _\eta^ \pm  } ,x_{\chi _\theta^ \pm  } ,x_{\tilde \nu_\sigma } ,x_{\tilde{D}_l^ -  } })
 C_L^{\tilde \nu_\sigma \mu \bar{\chi} _\eta^ \pm  } C_R^{\tilde \nu_\sigma \chi _\theta^ \pm  \bar e}
 C_R^{\tilde{D}_l^ - \chi _\eta^ \pm \bar{u}   } C_L^{\tilde{D}_l^ -  u \bar {\chi} _\theta^ \pm  } \nonumber\\
 &&\hspace{1.8cm} + \frac{{m_{\chi _\eta^ \pm  } m_{\chi _\theta^ \pm  } }}{{4e^2 m_W^4 }}E_8 ( {x_{\chi _\eta^ \pm  } ,x_{\chi _\theta^ \pm  } ,x_{\tilde \nu_\sigma } ,x_{\tilde{D}_l^ -  } })C_L^{\tilde \nu_\sigma \mu \bar{\chi} _\eta^ \pm  } C_R^{\tilde \nu_\sigma \chi _\theta^ \pm  \bar e}
 C_L^{\tilde{D}_l^ - \chi _\eta^ \pm \bar{u}   } C_R^{\tilde{D}_l^ -  u \bar {\chi} _\theta^ \pm   }  \Big], \nonumber\\
 &&O_q^R (c) = O_q^L (c)|_{L \leftrightarrow R} \; .
\end{eqnarray}

Furthermore, the virtual $W$ produces corrections through the diagrams  in the last line of Fig.\ref{FMFLV2}.
\begin{eqnarray}
 &&O_d^L (W) =  \sum_{\delta=1}^6- \frac{1}{{2e^2 }}\frac{\partial }{{\partial x_W }}E_2 ( {x_W ,x_{\nu_\delta}  ,x_u })
 C_L^{W\mu \bar {\nu}_\delta } C_L^{W^* {\nu}_\delta \bar e} C_L^{W^* u\bar d} C_L^{Wd\bar u}  , \nonumber\\
 &&O_u^L (W) = \sum_{\delta=1}^6\frac{2}{{e^2 }}\frac{\partial }{{\partial x_W }}E_2 ( {x_W ,x_{\nu_\delta}  ,x_d } )
 C_L^{W^* \mu \bar {\nu}_\delta } C_L^{W{\nu}_\delta \bar e} C_L^{W^* d\bar u} C_L^{Wu\bar d} , \nonumber\\
&&O_d^R (W) = O_u^R (W) = 0.
 \end{eqnarray}

Starting from the effective Lagrangian describing $\mu-e$ conversion processes at the quark level, one can determine the $\mu-e$ conversion rate within a nucleus~\cite{Bernabeu:1993ta}:
\begin{eqnarray}
&&{\rm{CR}}(\mu \to e:{\rm{Nucleus}}) \nonumber\\
&&\qquad = 4 \alpha^5 \frac{Z_{\rm{eff}}^4}{Z } \left| F(q^2) \right|^2 m_\mu^5  \Big[\left| Z( A_1^L  -  {A_2^R} ) - (2Z+N)\bar{O}_u^L - (Z+2N)\bar{O}_d^L \right| ^2 \nonumber\\
&&\qquad+ \: \left| Z( A_1^R  -  {A_2^L} ) - (2Z+N)\bar{O}_u^R - (Z+2N)\bar{O}_d^R \right|^2 \Big]\frac{1}{\Gamma_{\rm{capt}}},
\end{eqnarray}
with
\begin{eqnarray}
&&T_{3L}^u=\frac{1}{2},~T_{3L}^d=-\frac{1}{2},~T_{3R}^u=T_{3R}^d=0,~Q_{em}^u=\frac{2}{3},~Q_{em}^d=-\frac{1}{3}, \nonumber\\
&&Z_{L,R}^q = T_{3L,R}^q - Q_{em}^q s_{_W}^2,\quad (q=u,d),\nonumber\\
&&\bar{O}_q^L = O_q^L + \frac{Z_L^q+Z_R^q}{2} \frac{D_L}{{m_Z^2}s_{_W}^2c_{_W}^2},  \quad (s_{_W}=\sin\theta_{_W},~c_{_W}=\cos\theta_{_W}), \nonumber\\
&&\bar{O}_q^R  =  \left. {\bar{O}_q^L} \right|{ _{L \leftrightarrow R}},  \quad (q=u,d ). \quad
\end{eqnarray}

In this context, $Z$ and $N$ correspond to the number of protons and neutrons comprising a nucleus, whereas $Z_{\rm{eff}}$ denotes the effective atomic charge, as ascertained in the referenced studies \cite{Sens:1959zz,Zeff2}. $F(q^2)$ represents the nuclear form factor, and $\Gamma_{\rm{capt}}$ is the total muon capture rate. For an array of distinct nuclei, the respective values of $Z_{\rm{eff}}$, $F(q^2\simeq-m_\mu^2)$, and $\Gamma_{\rm{capt}}$ have been compiled in Table \ref{biao2}, in accordance with the methodology outlined in Ref. \cite{Kitano:2002mt}.

\begin{table}[h]
\caption{The values of $Z_{\rm{eff}}$, $F(q^2\simeq-m_\mu^2)$ and $\Gamma_{(\rm{capt})}$ for different nuclei}
\begin{tabular}{|c|c|c|c|c|}
\hline
$_{Z}^{A}{\rm{Nucleus}}$ & $Z_{\rm{eff}}$ & $F(q^2\simeq-m_\mu^2)$ & $\Gamma_{\rm{capt}}({\rm{GeV}})$ \\
\hline
$_{22}^{48}{\rm{Ti}}$ & 17.6 & 0.54 & $1.70422\times10^{-18}$ \\
\hline
$_{\:79}^{197}{\rm{Au}}$ & 33.5 & 0.16 & $8.59868\times10^{-18}$ \\
\hline
$_{\:82}^{207}{\rm{Pb}}$ & 34.0 & 0.15 & $8.84868\times10^{-18}$ \\
\hline
\end{tabular}
\label{biao2}
\end{table}

\section{numerical results}

In this section, we analyze the numerical results and consider the experimental constraints. In order to obtain reasonable numerical results, a number of sensitive parameters need to be investigated from those used. Given that experimental constrains from the $l_j\rightarrow{l_i\gamma}$ processes tightly constrain the parameter space of the N-B-LSSM, it is crucial to take into account the impacts of  $l_j\rightarrow{l_i\gamma}$ on LFV when analyzing $l_j \rightarrow 3l_i$ and $\mu\rightarrow e+q\bar q$ processes. If the strict conditions set by the $l_j\rightarrow{l_i\gamma}$ processes are satisfied, it is reasonable to expect that the constraints from other related LFV processes will also be satisfied.

Several experimental restrictions are considered, including:

1. The lightest CP-even Higgs $h^0$ mass is around 125.20 $\pm$ 0.11 GeV \cite{pdg,cms,atlas}.

2. According to the latest data from the LHC \cite{w1,w2,w3,w4,w5,w6}, the scalar lepton mass is greater than $700~{\rm GeV}$, the chargino mass is greater than $1100~{\rm GeV}$, and the scalar quark mass is greater than $1500~{\rm GeV}$.

3. The new angle $\tan{\beta}_{\eta}$ constrained by the LHC should be less than 1.5 \cite{lb}.

4. The $Z^\prime$ boson mass is larger than 5.1 TeV \cite{at}.

5. The ratio between $M_{Z^\prime}$ and its gauge $M_{Z^\prime}/g_B \geq 6 ~{\rm TeV}$ \cite{gc,ZPG2}.

6. The limitations from Charge and Color Breaking (CCB), neutrino experiment data and muon anomalous magnetic dipole moment are also taken into account \cite{HAN1,HAN2,Zhao:2015dna,L1,L2,L3}.

The Yukawa couplings of neutrinos $Y_{\nu ij}$ $(i,j=1,2,3)$ are at the order of $10^{-8}\sim10^{-6}$ and its effects on the LFV processes are small and usually negligible. Incorporating the experimental requirements, we compile and analyse extensive data, generating one-dimensional plots and scatter plots to illustrate the correlations between various parameters and branching ratios. Through the analysis of these plots in conjunction with the experimental constraints on branching ratios, a viable parameter space is identified to explain LFV.

Considering these limitations, we adopt the following parameters in the numerical calculation:
\begin{eqnarray}
&&\lambda_2=-0.25,~~~\kappa=0.1,~~\tan{\beta}_{\eta}=0.9,~~Y_{Xii}=0.5,\nonumber\\
&&T_{\kappa} =-2.5~ {\rm TeV},~~T_{\lambda}=T_{\lambda_2}=1~{\rm TeV},
~~M_1=0.4~{\rm TeV},\nonumber\\
&&M_{BL}=1~{\rm TeV},~~M_{BB'}=0.1~{\rm TeV},~~M_{\nu ii}^2=2.5~{\rm TeV}^2,\nonumber\\
&&T_{uii}=1~{\rm TeV},~~T_{dii}=1~{\rm TeV},~~T_{Xii}=-4~{\rm TeV},(i=1,2,3).
\end{eqnarray}

To facilitate the numerical investigation, we employ the parameter relationships and analyze their variations in the subsequent numerical analysis:
\begin{eqnarray}
&&\tan\beta,~~g_B,~~g_{YB},~~\lambda,~~v_S,~~M_2,~~M_{\nu ii}^2=M_{\nu}^2,~~M_{Lii}^2=M_{L}^2,\nonumber\\
&&M_{Eii}^2=M_{E}^2,~~M_{\nu ij}^2=M_{\nu ji}^2,~~M_{Lij}^2=M_{Lji}^2,~~M_{Eij}^2=M_{Eji}^2,\nonumber\\
&&T_{eii}=T_{e},~~T_{\nu ii}=T_{\nu},~~T_{eij}=T_{eji},~~T_{\nu ij}=T_{\nu ji},(i,j=1,2,3,~i\neq j).
\end{eqnarray}

If no special case exists, the non-diagonal elements of the parameters are assumed to be zero.

\subsection{$l_j\rightarrow l_i\gamma$}

LFV is closely associated with NP, and the branching ratio of the process $\mu\rightarrow e\gamma$ is subject to stringent experimental constraints. The latest experimental upper limit on its branching ratio is $4.2\times10^{-13}$ at $90\%$ confidence level. At this subsection, the chosen parameters are $M_2=1.2~{\rm TeV}, v_S=4~{\rm TeV}, \lambda=0.4, T_\nu=1~{\rm TeV}, M_{E}^2=1.5~{\rm TeV}^2$.

1. $\mu\rightarrow e\gamma$

\begin{figure}[ht]
\setlength{\unitlength}{5mm}
\centering
\includegraphics[width=2.9in]{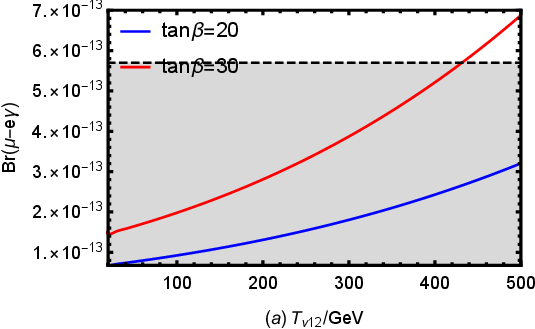}
\setlength{\unitlength}{5mm}
\centering
\includegraphics[width=2.9in]{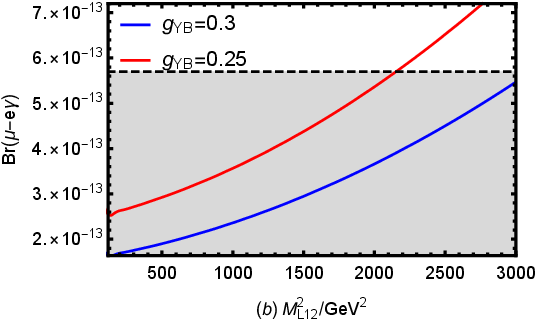}
\setlength{\unitlength}{5mm}
\centering\nonumber\\
\includegraphics[width=2.9in]{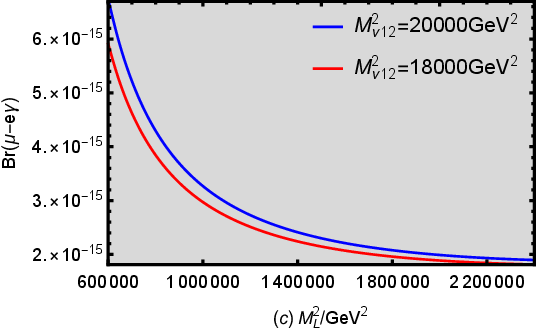}
\setlength{\unitlength}{5mm}
\centering
\includegraphics[width=2.9in]{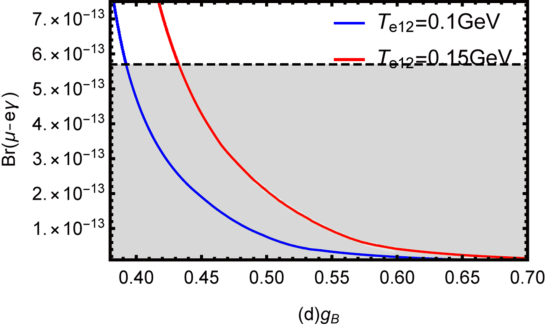}
\caption{Br($\mu\rightarrow e\gamma$) schematic diagrams affected by different parameters, the gray area satisfies the experimental upper limit. Fig.\ref{mue}(a) shows the relationship between $T_{\nu12}$ and Br($\mu\rightarrow e\gamma$) with the red line representing $\tan\beta$=30 and the blue line representing $\tan\beta$=20. Fig.\ref{mue}(b) shows the relationship between $M_{L12}^2$ and Br($\mu\rightarrow e\gamma$) with the red line representing $g_{YB}=0.25$ and the blue line representing $g_{YB}=0.3$. Fig.\ref{mue}(c) shows the relationship between $M_{L}^2$ and Br($\mu\rightarrow e\gamma$) with the red line representing $M_{\nu12}^2=1.8\times10^{4}~{\rm GeV}^2$ and the blue line representing $M_{L12}^2=2\times10^{4}~{\rm GeV}^2$. Fig.\ref{mue}(d) shows the relationship between $g_B$ and Br($\mu\rightarrow e\gamma$) with the red line representing $T_{e12}=0.15~{\rm GeV}$ and the blue line representing $T_{e12}=0.1~{\rm GeV}$.}\label{mue}
\end{figure}

We conduct numerical calculations for Br($\mu\rightarrow e\gamma$), and depict the impacts of various parameters in Fig.\ref{mue}. The gray area indicates the region satisfying the experimental upper limit.

With the parameters $g_{YB}=0.2,~g_B=0.6,~T_e=1~{\rm TeV},~M_{L}^2=0.16~{\rm TeV}^2$, the variation of Br($\mu\rightarrow e\gamma$) with respect to $T_{\nu12}$ is depicted in Fig.\ref{mue}(a). The red line corresponds to $\tan\beta$=30 and the blue line corresponds to $\tan\beta$=20. $\tan\beta$ represents the ratio of the VEVs of two Higgs doublets ($\tan\beta=\upsilon_{u} / \upsilon_{d}$), and it influences particle masses by directly affecting $v_u$ and $v_d$. $\tan\beta$ is closely associated with the mass matrixes of chargino, neutralino, slepton and sneutrino, particularly their non-diagonal elements. In the N-B-LSSM, nearly all contributions to LFV processes are impacted by $\tan\beta$. As a highly sensitive parameter, $\tan\beta$ has a significant effect on numerical results. It can be observed from Fig.\ref{mue}(a) that Br($\mu\rightarrow e\gamma$) rises as $\tan\beta$ increases. As a non-diagonal element of the sneutrino mass matrix, $T_{\nu12}$ influences the masses and mixing of sneutrino. In the range from 20 {\rm GeV}-500 {\rm GeV}, Br($\mu\rightarrow e\gamma$) increases with the augment of $T_{\nu12}$. Both $\tan\beta$ and $T_{\nu12}$ are sensitive parameters, and since the upper limit for  Br($\mu\rightarrow e\gamma$) is small, it is easy to exceed that upper limit in the N-B-LSSM.

Supposing $\tan\beta=50,~g_B=0.6,~T_e=1~{\rm TeV},~M_{L}^2=0.16~{\rm TeV}^2$, we represent the variation of Br($\mu\rightarrow e\gamma$) with $M_{L12}^2$ using red line ($g_{YB}=0.25$) and blue line ($g_{YB}=0.3$) in Fig.\ref{mue}(b). $g_{YB}$ is the mixing gauge coupling constant of $U(1)_Y$ group and $U(1)_{B-L}$ group, it describes the interactions and mixing effects between two $U(1)$ gauge groups, which is a new parameter that goes beyond MSSM and can bring new effects. $M_{L12}^2$ is the flavour mixing parameter that appears in the mass matrixes of the slepton, CP-even sneutrino and CP-odd sneutrino. We can clearly see that these two lines increase with $M_{L12}^2$. Overall the red line is larger than the blue line, the blue line as a whole and the red line in the range 120 ${\rm GeV}^2$ $<M_{L12}^2<2100$ ${\rm GeV}^2$ are located in the gray area that satisfies the limits of the experiment, and the rest of the red line goes beyond the gray area.

Based on $\tan\beta=25,~g_B=0.3,~g_{YB}=0.1,~T_e=1~{\rm TeV}$, the blue line in Fig.\ref{mue}(c) corresponds to $M_{\nu12}^2=2\times10^{4}~{\rm GeV}^2$ and the red line corresponds to $M_{\nu12}^2=1.8\times10^{4}~{\rm GeV}^2$. These two lines show the trend of Br($\mu\rightarrow e\gamma$) with the change of $M_{L}^2$ at different values of $M_{\nu12}^2$. As $M_{L}^2$ increases, both lines show a gradual decrease in Br($\mu\rightarrow e\gamma$), and the two lines are very close to each other in the whole $M_{L}^2$ range. The similarity in the trends could suggest that $M_{L}^2$ exhibits greater sensitivity compared to $M_{\nu12}^2$, this is because $M_{L}^2$ impacts the masses of both slepton and sneutrino, while $M_{\nu12}^2$ only influences sneutrino masses.

Setting $\tan\beta=25,~g_{YB}=0.1,~T_e=1~{\rm TeV},~M_{L}^2=0.16~{\rm TeV}^2$, we study the effect of the parameter $g_B$ on Br($\mu\rightarrow e\gamma$) in Fig.\ref{mue}(d), where the red and blue lines represent the cases of $T_{e 12}=0.15~{\rm GeV}$ and $T_{e 12}=0.1~{\rm GeV}$, respectively. $g_B$ is the coupling constant of the gauge group. $g_B$ appears in almost all mass matrixes and determines how bosons interact with fermions, Higgs field and other particles. So it is clear that $g_B$ is a sensitive parameter. As depicted in Fig.\ref{mue}(d), the value of Br($\mu\rightarrow e\gamma$) decreases with an increase in parameter $g_B$. There is $T_{e 12}$ as a non-diagonal element in the slepton mass squared matrix, which influences the results through neutralino-slepton contributions. It is evident that despite a change of only $0.05~{\rm GeV}$ in $T_{e 12}$, it has led to a significant difference in Br($\mu\rightarrow e\gamma$). Specifically, when $T_{e 12}$ increase from $0.1~{\rm GeV}$ to $0.15~{\rm GeV}$,  Br($\mu\rightarrow e\gamma$) increases for the same value of $g_B$.

2. $\tau\rightarrow \mu\gamma$

\begin{figure}[ht]
\setlength{\unitlength}{5mm}
\centering
\includegraphics[width=2.9in]{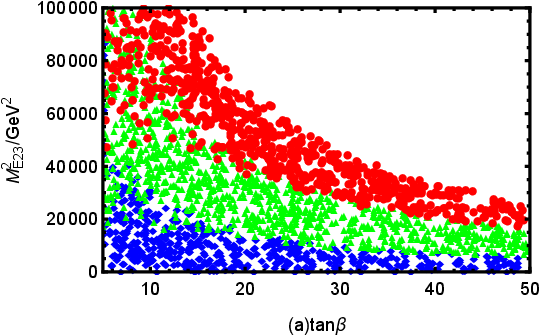}
\setlength{\unitlength}{5mm}
\centering
\includegraphics[width=2.9in]{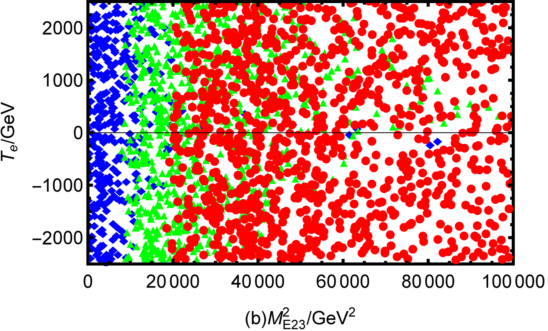}
\caption{Under the constraint of Br($\tau\rightarrow \mu\gamma$) process, a reasonable parameter space is selected for random scattering, and the marking of the scattering points represents: $\textcolor{blue}{\blacklozenge}~(0<Br(\tau\rightarrow \mu\gamma)<2\times 10^{-9}),~\textcolor{green}{\blacktriangle}~(2\times 10^{-9}\leq Br(\tau\rightarrow \mu\gamma)<1.7\times 10^{-8}),~ \textcolor{red}{\bullet}~(1.7\times 10^{-8}\leq Br(\tau\rightarrow \mu\gamma)<4.4\times 10^{-8})$.}{\label {taomu}}
\end{figure}

The upper bound of Br($\tau\rightarrow \mu\gamma$) experiment is $4.2\times10^{-8}$, which is nearly 5 orders of magnitude lager than the experimental upper bound of Br($\mu\rightarrow e\gamma$). First, we set the parameters $g_{YB}=0.1,~g_B=0.3,~M_{L}^2=0.16~{\rm TeV}^2$. Next, we perform random scattering with parameters and ranges shown in Table \ref{biao3}. Among these parameters, there are both sensitive and insensitive parameters. Based on the randomly scanned data, we draw Fig.\ref{taomu} to find the rules. In order to observe the numerical patterns more clearly and to facilitate the analysis of the results, we divide the numerical data obtained from the random scattering points into different ranges. We use $\textcolor{blue}{\blacklozenge}~(0<Br(\tau\rightarrow \mu\gamma)<2\times 10^{-9}),~\textcolor{green}{\blacktriangle}~(2\times 10^{-9}\leq Br(\tau\rightarrow \mu\gamma<1.7\times 10^{-8}),~ \textcolor{red}{\bullet}~(1.7\times 10^{-8}\leq Br(\tau\rightarrow \mu\gamma<4.4\times 10^{-8})$ to denote the results.

The relationship between $\tan\beta$ and $M^2_{E23}$ is shown in Fig.\ref{taomu}(a). $M^2_{E23}$ is a non-diagonal element that appears in the slepton mass matrix and affects the numerical results by influencing the slepton mixing and masses. Fig.\ref{taomu}(a) as a whole presents a triangle shape gradually shrinking from the top left to the bottom right, with data points densely distributed in the lower left corner and almost no points in the upper right corner. $\textcolor{blue}{\blacklozenge}$ are mainly distributed in the region of $M^2_{E23}<2\times10^4~\rm GeV^2$ and cover the whole range of $\tan\beta$ (from 5 to 50), with a relatively even distribution and no obviously downward trend. The $M^2_{E23}$ distribution of $\textcolor{green}{\blacktriangle}$ ranges from about $1\times10^4~\rm GeV^2$ to $4\times10^4~\rm GeV^2$. $\textcolor{green}{\blacktriangle}$ density gradually decreases as $\tan\beta$ increases, suggesting that fewer data satisfy this range for larger value of $\tan\beta$. $\textcolor{red}{\bullet}$ are concentrated in the high $M^2_{E23}$ region, ranging from about $2\times10^4~\rm GeV^2$ to $1\times10^5~\rm GeV^2$. The number of $\textcolor{red}{\bullet}$ decreases rapidly with the increase of $\tan\beta$, and especially after $\tan\beta>20$. The maximum value of $M^2_{E23}$ decreases significantly, showing a steep negative correlation. The reason for this trend may be due to the existence of a non-linear relationship between the parameters, where an increase in $\tan\beta$ imposes a stronger restriction on $M_{E23}^2$, resulting in a rapid decrease in the value of $M_{E23}^2$, which leads to a gradual convergence of the data points into the lower $M_{E23}^2$ region, and in addition Br($\tau\rightarrow \mu\gamma$) imposes tighter constraints on the parameter space, prohibiting the co-existence of high $\tan\beta$ and high $M_{E23}^2$.

\begin{table*}
\caption{Scanning parameters for Fig.{\ref {taomu}}}\label{biao3}
\begin{tabular}{|c|c|c|}
\hline
Parameters&Min&Max\\
\hline
$\hspace{1.5cm}M^2_{\nu}/\rm GeV^2\hspace{1.5cm}$ &$\hspace{1.5cm}1\times10^{5}\hspace{1.5cm}$& $\hspace{1.5cm}2\times10^{6}\hspace{1.5cm}$\\
\hline
$\hspace{1.5cm}M^2_{E23}/\rm GeV^2\hspace{1.5cm}$ &$\hspace{1.5cm}0\hspace{1.5cm}$& $\hspace{1.5cm}10^5\hspace{1.5cm}$\\
\hline
$\hspace{1.5cm}T_{e}/\rm GeV\hspace{1.5cm}$ &$\hspace{1.5cm}-2500\hspace{1.5cm}$ &$\hspace{1.5cm}2500\hspace{1.5cm}$\\
\hline
$T_{\nu 23}$/GeV & $\hspace{1.5cm}-500\hspace{1.5cm}$ &$\hspace{1.5cm}500\hspace{1.5cm}$\\
\hline
$\tan\beta$ &$\hspace{1.5cm}5\hspace{1.5cm}$ &$\hspace{1.5cm}50\hspace{1.5cm}$\\
\hline
$g_B$ &$\hspace{1.5cm}0.3\hspace{1.5cm}$ &$\hspace{1.5cm}0.6\hspace{1.5cm}$\\
\hline
\end{tabular}
\end{table*}

The variation of $T_e$ with $M^2_{E23}$ is plotted in Fig.\ref{taomu}(b). $\textcolor{blue}{\blacklozenge}$ are concentrated in the low $M^2_{E23}$, $\textcolor{green}{\blacktriangle}$ are distributed in the middle region, and $\textcolor{red}{\bullet}$ are concentrated in the high $M^2_{E23}$ region, showing a trend of distribution from left to right, and the range of data points expands gradually with the change of color. Point of all colors are relatively uniformly distributed over $T_e$, with a range that almost completely cover from $-2500\rm GeV$ to $2500\rm GeV$. $T_e$ is also related to the slepton mass matrix, and we can analyse that $T_e$ does not strongly influence the Br($\tau\rightarrow \mu\gamma$) compared to $M^2_{E23}$. In summary, Fig.\ref{taomu}(b) is symmetric about $T_e=0~\rm GeV$ and the value of Br($\tau\rightarrow \mu\gamma$) becomes lager when the non-diagonal element $M^2_{E23}$ as the flavor mixing parameter is increased.

3. $\tau\rightarrow e\gamma$

\begin{figure}[ht]
\setlength{\unitlength}{5mm}
\centering
\includegraphics[width=2.7in]{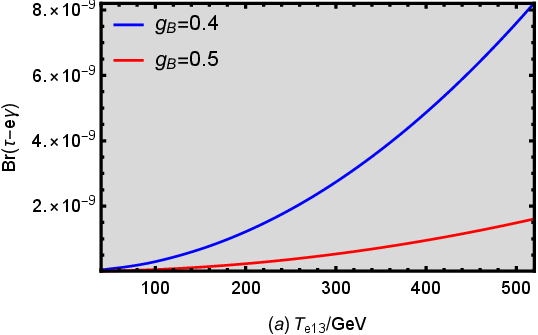}
\setlength{\unitlength}{5mm}
\centering
\includegraphics[width=2.9in]{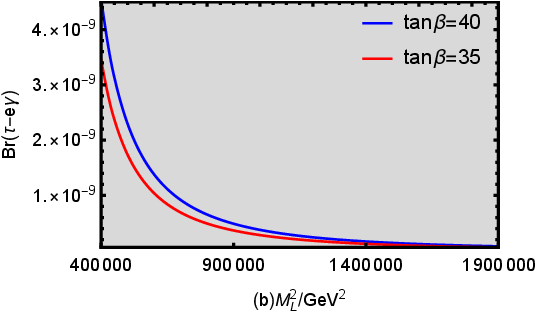}
\caption{Br($\tau\rightarrow e\gamma$) schematic diagrams affected by different parameters, the gray area satisfies the experimental upper limit. Fig.\ref{taoe}(a) shows the relationship between $T_{e 13}$ and Br($\tau\rightarrow e\gamma$) with the red line representing $g_B$=0.5 and the blue line representing $g_B$=0.4. Fig.\ref{taoe}(b) shows the relationship between $M_{L}^2$ and Br($\tau\rightarrow e\gamma$) with the red line representing $\tan\beta=35$ and the blue line representing $\tan\beta=40$.}{\label {taoe}}
\end{figure}

Similar to $\tau\rightarrow \mu\gamma$, $\tau\rightarrow e\gamma$ also has a large branching ratio, and the experimental upper bound is $3.3\times10^{-8}$. We study the influence of $T_{e 13},~g_B,~M^2_{L}$ and $\tan\beta$ on Br($\tau\rightarrow e\gamma$) in Fig.\ref{taoe}.

Using the parameters $g_{YB}=0.1,~\tan\beta=25,~T_e=1~{\rm TeV},~M_{L}^2=0.16~{\rm TeV}^2$, Fig.\ref{taoe}(a) shows the relationship between Br($\tau\rightarrow e\gamma$) and $T_{e 13}$ for the two scenarios $g_B=0.4$ (the blue line) and $g_B=0.5$ (the red line), respectively. The parameter $T_{e 13}$ refers to one of the trilinear terms related to the soft supersymmetry breaking associated with the lepton Yukawa coupling. Whether $g_B=0.4$ or $g_B=0.5$, Br($\tau\rightarrow e\gamma$) increases with $T_{e 13}$, and the increase is very significant over large $T_{e 13}$ ranges. To be specific, Br($\tau\rightarrow e\gamma$) increases faster with $T_{e 13}$ for $g_B=0.4$. When $T_{e 13}$ is small (less than $100~\rm GeV$), the value of Br($\tau\rightarrow e\gamma$) is very low and close to zero. As $T_{e 13}$ exceeds $100~\rm GeV$, Br($\tau\rightarrow e\gamma$) begins to grow significantly. In contrast, when $g_B=0.5$, the growth rate of Br($\tau\rightarrow e\gamma$) is significantly slower than that for $g_B=0.4$. It can be concluded that the size of $g_B$ affects the growth rate of the branching ratio. A reduction in the parameter $g_B$ leads to a diminished contribution from the diagonal matrix elements, thereby indirectly enhancing the relative influence of the non-diagonal terms. Conversely, a larger $g_B$ suppresses this growth effect, keeping the branching ratio well below the experimental upper limit.

Let $g_{YB}=0.1,~g_B=0.3,~T_e=1~{\rm TeV},~T_{\nu ij}=0.5~{\rm TeV},~M_{Lij}^2=0.016~{\rm TeV}^2$ ($i,j=1,2,3,~i\neq j$) in Fig.\ref{taoe}(b), the trend of Br($\tau\rightarrow e\gamma$) with $M_{L}^2$ is shown and the effect of $\tan\beta=35$ (rad line) and $\tan\beta=40$ (blue line) on the branching ratio is compared. As $M_{L}^2$ increases, Br($\tau\rightarrow e\gamma$) decreases significantly and gradually tends to zero, with the decrease being more pronounced in the low $M_{L}^2$ region $(4\times 10^{5}~{\rm GeV}^2\leq M_{L}^2<9\times 10^{5}~{\rm GeV}^2)$ and flattening out at high $M_{L}^2$ region $(9\times 10^{5}~{\rm GeV}^2\leq M_{L}^2<1.9\times 10^{6}~{\rm GeV}^2)$. An increasing in $M_{L}^2$ directly suppresses the amplitude of the lepton transition process, because $M_{L}^2$ presents in the diagonal elements of the slepton and sneutrino mass matrices. As $M_{L}^2$ increases, the mass eigenvalues of the particles also increase, and these mass eigenvalues appear in the denominators of the propagators. Larger mass eigenvalues weaken the transition amplitudes, leading to a decrease in the branching ratio. The branching ratio at $\tan\beta=40$ is consistently higher than that at $\tan\beta=35$ for the same $M_{L}^2$, and this difference is more obvious in the smaller $M_{L}^2$ region.

\subsection{$l_j\rightarrow 3l_i$}

In this subsection, we numerically investigate the LFV processes $l_j\rightarrow 3l_i$ under the assumption of $T_e=1~{\rm TeV}$. These processes are closely related to $l_j\rightarrow l_i\gamma$. Line graphs and scatter plots are drawn from the data obtained.

1. $\mu\rightarrow 3e$

\begin{figure}[ht]
\setlength{\unitlength}{5mm}
\centering
\includegraphics[width=2.9in]{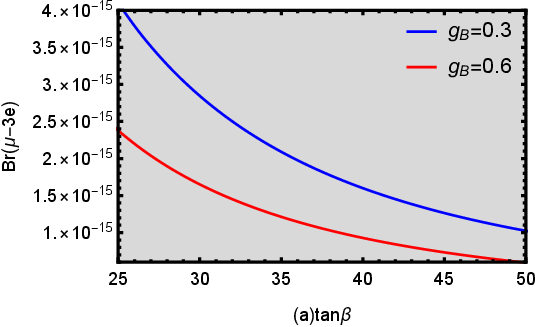}
\setlength{\unitlength}{5mm}
\centering
\includegraphics[width=2.9in]{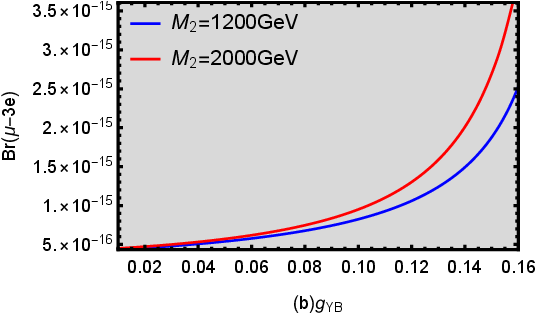}
\setlength{\unitlength}{5mm}
\centering\nonumber\\
\includegraphics[width=2.9in]{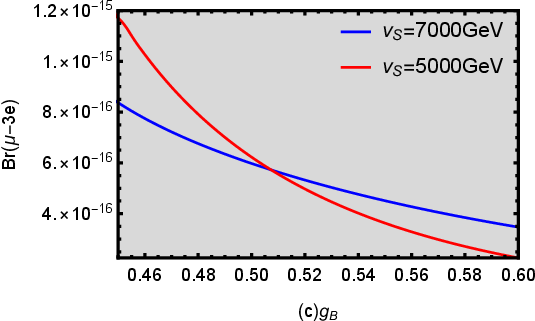}
\setlength{\unitlength}{5mm}
\centering
\includegraphics[width=2.9in]{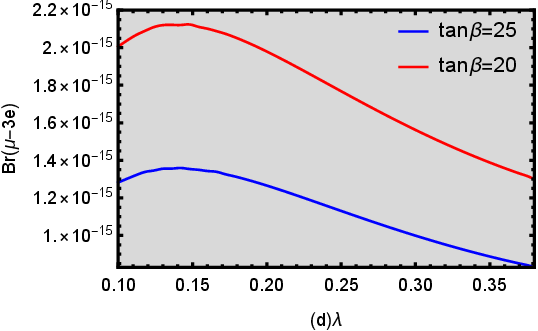}
\caption{Br($\mu\rightarrow 3e$) schematic diagrams affected by different parameters, the gray area satisfies the experimental upper limit. Fig.\ref{mu3e}(a) shows the relationship between $\tan\beta$ and Br($\mu\rightarrow 3e$) with the red line representing $g_B$=0.6 and the blue line representing $g_B$=0.3. Fig.\ref{mu3e}(b) shows the relationship between $g_{YB}$ and Br($\mu\rightarrow 3e$) with the red line representing $M_2=2000~{\rm GeV}$ and the blue line representing $M_2=1200~{\rm GeV}$. Fig.\ref{mu3e}(c) shows the relationship between $g_B$ and Br($\mu\rightarrow 3e$) with the red line representing $v_S=5000~{\rm GeV}$ and the blue line representing $v_S=7000~{\rm GeV}$. Fig.\ref{mue}(d) shows the relationship between $\lambda$ and Br($\mu\rightarrow 3e$) with the red line representing $\tan\beta=20$ and the blue line representing $\tan\beta=25$.}{\label {mu3e}}
\end{figure}

Br$(\mu\rightarrow 3e)$ is the strictest branching ratio of LFV processes $l_j\rightarrow 3l_i$ and the experiment upper bound is $1.0 \times 10^{-12}$. In order to clearly find the sensitive parameters affecting $\mu\rightarrow 3e$, we plot Fig.\ref{mu3e} for different parameters.

We set the parameters $g_{YB}=0.02,~\lambda=0.4,~M_2=1.2~{\rm TeV},~v_S=4~{\rm TeV},~T_\nu=1~{\rm TeV},~T_{e 12}=3\times10^{-4}~{\rm TeV}, ~M_{L}^2=0.16~{\rm TeV}^2,~M_{E}^2=1.5~{\rm TeV}^2$ in Fig.\ref{mu3e}(a). It shows the variation of Br$(\mu\rightarrow 3e)$ with $\tan\beta$, where the lines correspond to different $g_B$ values (the blue line corresponds to $g_B$=0.3, the red line corresponds to $g_B$=0.6). It can be observed that Br$(\mu\rightarrow 3e)$ gradually decreases as $\tan\beta$ increases. Moreover, for the same value of $\tan\beta$, Br$(\mu\rightarrow 3e)$ at $g_B$=0.6 is smaller than that at $g_B$=0.3, this indicates that an increase in the value of $g_B$ also leads to a decrease in Br$(\mu\rightarrow 3e)$.

In the case of $\tan\beta=25,~g_{B}=0.3,~\lambda=0.4,~v_S=4~{\rm TeV},~T_\nu=1~{\rm TeV},~T_{e 12}=1\times10^{-4}~{\rm TeV}, ~M_{L}^2=0.16~{\rm TeV}^2,~M_{E}^2=1.5~{\rm TeV}^2$, we plot Br$(\mu\rightarrow 3e)$ varying with $g_{YB}$ in Fig.\ref{mu3e}(b), where the blue line is $M_2=1200~{\rm GeV}$ and the red line is $M_2=2000~{\rm GeV}$. $M_2$ affects the mass matrixes of the neutralino and Chargino. At the same $g_{YB}$, the branching ratio corresponding to the larger $M_2$ is significantly higher than that of the smaller $M_2$. $g_{YB}$ is a sensitive parameter, Br$(\mu\rightarrow 3e)$ rises rapidly with $g_{YB}$. This trend reflects the fact that $g_{YB}$ enhances the interaction strength of LFV process, while the increase of $M_2$ attenuates the inhibitory effect of the new physics particles masses on the propagation process, both of which together lead to the branching ratio growing in a nonlinear manner.

$\tan\beta=25,~g_{YB}=0.1,~\lambda=0.4,~M_2=1.2~{\rm TeV},~T_\nu=1~{\rm TeV},~T_{e 12}=1\times10^{-4}~{\rm TeV},~M_{L}^2=0.16~{\rm TeV}^2,  ~M_{E}^2=1.5~{\rm TeV}^2$ are set to study the relationship between Br$(\mu\rightarrow 3e)$ and $g_{B}$ in Fig.\ref{mu3e}(c). The lines are divided into two cases corresponding to $v_S=5000~{\rm GeV}$ (red line) and $v_S=7000~{\rm GeV}$ (blue line). $v_S$ is the VEV of $S$ and appears in the diagonal elements of CP-even Higgs mass squared matrix. More importantly, $v_S$ affects the mass of the lightest neutralino through $m_{{\tilde{\chi}_1}{\tilde{\chi}_2}}=-\frac{1}{\sqrt{2}} {\lambda_2} v_S$. In the $g_B<0.5$ region, the red line is higher than the blue line. In the $g_B>0.5$ region, the blue line is higher than the red line. Near $g_B=0.5$, the two lines cross, which means that the branching ratios of the two $v_S$ parameters are equal for a given value of $g_B$. $g_B$ is the coupling constant for the new physical interactions, and for two different values of $v_S$, Br$(\mu\rightarrow 3e)$ both show a monotonically decreasing trend with increasing $g_B$.

Defining the parameters $\tan\beta=25,~g_{YB}=0.1,~g_{B}=0.3,~v_S=4~{\rm TeV},~M_2=1.2~{\rm TeV},~T_\nu=1~{\rm TeV},~T_{e 12}=1\times10^{-4}~{\rm TeV}, ~M_{L}^2=0.16~{\rm TeV}^2,~M_{E}^2=1.5~{\rm TeV}^2$. Fig.\ref{mu3e}(d) shows the variation trend of Br$(\mu\rightarrow 3e)$ with $\lambda$ under the values of two different parameter $\tan\beta$, where the blue line represents $\tan\beta=25$ and the red line represents $\tan\beta=20$. $\lambda$ is the constant of the $\lambda\hat{S}\hat{H}_u\hat{H}_d$ term in the superpotential and appears in the mass squared matrix of neutralino at tree level. Both lines show a tendency for Br$(\mu\rightarrow 3e)$ to increase slightly and then gradually decrease as $\lambda$ increases, with Br$(\mu\rightarrow 3e)$ reaching a maximum near the value of $\lambda$=0.15. The height of the peak varies with $\tan\beta$, but the overall shape is similar. The Br$(\mu\rightarrow 3e)$ value at $\tan\beta$=20 is always higher than the value at $\tan\beta$=25.

2. $\tau\rightarrow 3\mu$

The experiment upper bound of the LFV process Br$(\tau\rightarrow 3\mu)$ is $2.1\times10^{-8}$, which is four orders of magnitude larger than that of $\mu\rightarrow 3e$. Here it is assumed that $g_{YB}=0.1,~g_B=0.3,~\lambda=0.4,~v_S=4~{\rm TeV},~M_2=1.2~{\rm TeV},~M_{E}^2=1.5~{\rm TeV}^2$ to study $\tau\rightarrow 3\mu$.

\begin{figure}[ht]
\setlength{\unitlength}{5mm}
\centering
\includegraphics[width=2.9in]{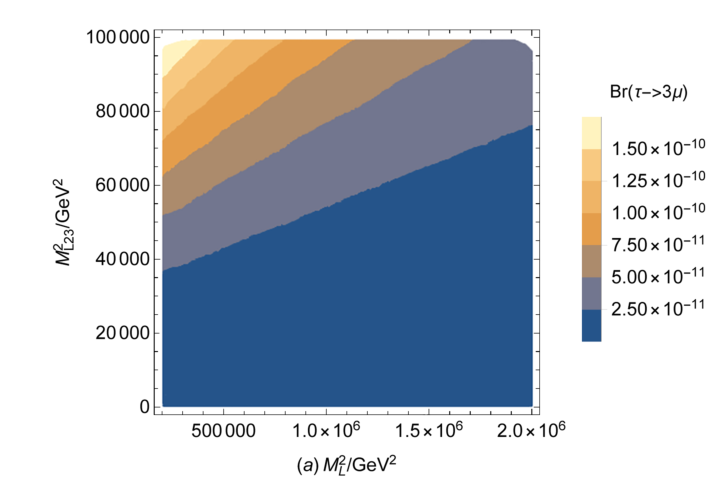}
\setlength{\unitlength}{5mm}
\centering
\includegraphics[width=2.7in]{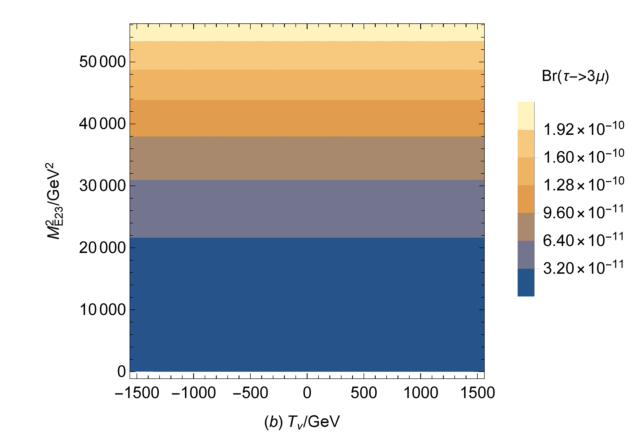}
\setlength{\unitlength}{5mm}
\centering\nonumber\\
\includegraphics[width=2.9in]{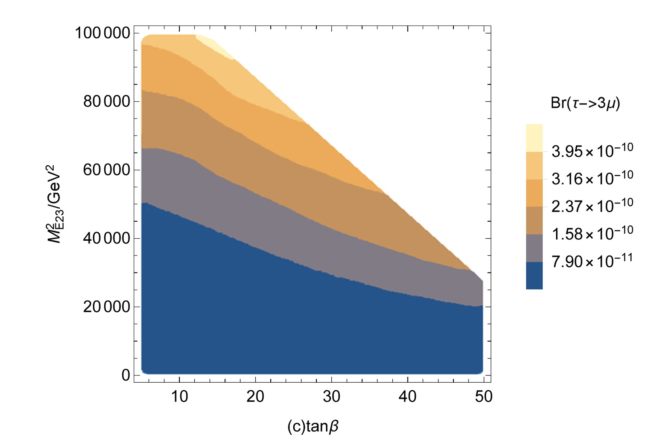}
\caption{(a) shows the relationship of $M_{L}^2$ and $M_{L23}^2$ on Br$(\tau\rightarrow 3\mu)$. The horizontal coordinate indicates the range $2\times10^{5}~{\rm GeV^2}\leq M_{L}^2\leq2\times10^{6}~{\rm GeV^2}$ and the vertical coordinate indicates the range $0~{\rm GeV^2}\leq M_{L23}^2\leq1\times10^{5}~{\rm GeV^2}$. (b) shows the relationship of $T_\nu$ and $M_{E23}^2$ on Br$(\tau\rightarrow 3\mu)$. The horizontal coordinate indicates the range $-1500~{\rm GeV}\leq T_{\nu}\leq1500~{\rm GeV}$ and the vertical coordinate indicates the range $0~{\rm GeV^2}\leq M_{E23}^2\leq55000~{\rm GeV^2}$. (c)shows the relationship of $\tan\beta$ and $M_{E23}^2$ on Br$(\tau\rightarrow 3\mu)$. The horizontal coordinate indicates the range $5\leq \tan\beta \leq50$ and the vertical coordinate indicates the range $0~{\rm GeV^2}\leq M_{E23}^2\leq1\times10^{5}~{\rm GeV^2}$. The right icons represent the colors corresponding to the the Br$(\tau\rightarrow 3\mu)$ values.}{\label {tao3mu}}
\end{figure}

In order to explore the influence of $M_{L}^2,~M_{L23}^2$ on $\tau\rightarrow 3\mu$ in Fig.\ref{tao3mu}(a), we set $T_\nu=1~{\rm TeV},~\tan\beta=25$ and perform random scans in the following ranges:
\begin{eqnarray}
&&0.02\leq g_{YB}\leq0.3,~~2\times10^{5}~{\rm GeV^2}\leq M_{L}^2\leq2\times10^{6}~{\rm GeV^2},\nonumber\\&&-2500~{\rm GeV}\leq T_{\nu}\leq2500~{\rm GeV},~~0~{\rm GeV^2}\leq M_{L23}^2\leq1\times10^{5}~{\rm GeV^2}.
\end{eqnarray}

As can be seen from Fig.\ref{tao3mu}(a), the variation of Br$(\tau\rightarrow 3\mu)$ shows a significant dependence. When $M_{L23}^2$ is small (in the region of vertical coordinate $M_{L23}^2 <40000~{\rm GeV^2}$), both the changes of $M_{L}^2$ and $M_{L23}^2$ have a weaker effect on the branching ratios, resulting in insensitive changes in branching ratios. This is due to the small contribution of the non-diagonal element $M_{L23}^2$ to the flavor mixing, while the mass suppression effect of $M_{L}^2$ has suppressed the branching ratios to a low level. At larger $M_{L23}^2$ (in the region of vertical coordinates $M_{L23}^2>40000~{\rm GeV^2}$), the dependence of branching ratios on both becomes more pronounced: branching ratios decrease rapidly with increasing $M_{L}^2$, reflecting the dominant effect of mass suppression effects; meanwhile, $M_{L23}^2$ increase also significantly increases branching ratios, reflecting the contribution of non-diagonal element to enhance flavor mixing. Near the contour line (color boundary), the change in branching ratio presents as a diagonal line, suggesting that an increase in $M_{L23}^2$ partially compensates for the mass suppression effect of $M_{L}^2$. This suggests that as the non-diagonal element get larger and the diagonal element get smaller, the branching ratio of the $(\tau\rightarrow 3\mu)$ process is closer to the experimental upper limit.

In Fig.\ref{tao3mu}(b), we set $M_{L}^2=1.6\times10^{5}~{\rm GeV^2},~\tan\beta=25$ to study the effect the two parameters $T_\nu$ and $M_{E23}^2$ together on Br$(\tau\rightarrow 3\mu)$. The ranges of scattering points are as follows:
\begin{eqnarray}
&&-2500~{\rm GeV}\leq T_{\nu}\leq2500~{\rm GeV},~~2\times10^{5}~{\rm GeV^2}\leq M_{L}^2\leq2\times10^{6}~{\rm GeV^2},\nonumber\\&&-500~{\rm GeV}\leq T_{e 23}\leq500~{\rm GeV},~~0~{\rm GeV^2}\leq M_{E23}^2\leq1\times10^{5}~{\rm GeV^2}.
\end{eqnarray}

It can be seen from Fig.\ref{tao3mu}(b) that Br$(\tau\rightarrow 3\mu)$ hardly changes with the horizontal coordinate $T_\nu$, but increases completely with the vertical coordinate $M_{E23}^2$. $T_\nu$ is the parameter related to the non-diagonal element of the sneutrino mass squared matrix, and it follows that in this parameter space Br$(\tau\rightarrow 3\mu)$ dependence on $T_\nu$ is negligible, and the influence of $M_{E23}^2$ on Br$(\tau\rightarrow 3\mu)$ predominates.

Let us assume that $M_{L}^2=1.6\times10^{5}{\rm GeV^2},~T_\nu=1~{\rm TeV}$, we focus on the effect of $\tan\beta$ and $M_{E23}^2$ on Br$(\tau\rightarrow 3\mu)$ in Fig.\ref{tao3mu}(c). We randomly scan the parameters as follows:
\begin{eqnarray}
&&5\leq \tan\beta\leq50,~~2\times10^{5}~{\rm GeV^2}\leq M_{L}^2\leq2\times10^{6}~{\rm GeV^2},\nonumber\\&&-2500~{\rm GeV}\leq T_{\nu}\leq2500~{\rm GeV},~~0~{\rm GeV^2}\leq M_{E23}^2\leq1\times10^{5}~{\rm GeV^2}.
\end{eqnarray}

Fig.\ref{tao3mu}(c) presents a trapezoidal distribution as a whole, and $M_{E23}^2$ is a non-diagonal element of slepton mass matrix. The increase of $M_{E23}^2$ will significantly enhance slepton flavor mixing effect, which brings the same impact as Fig.\ref{tao3mu}(b). Increasing $M_{E23}^2$ significantly increases Br$(\tau\rightarrow 3\mu)$, and increasing $\tan\beta$ also increases Br$(\tau\rightarrow 3\mu)$, but its effect is limited in scope. The white area in the upper right corner of Fig.\ref{tao3mu}(c) may be due to the limitation of the masses of the Higgs particle and other particles.

3. $\tau\rightarrow 3e$

\begin{figure}[ht]
\setlength{\unitlength}{5mm}
\centering
\includegraphics[width=2.9in]{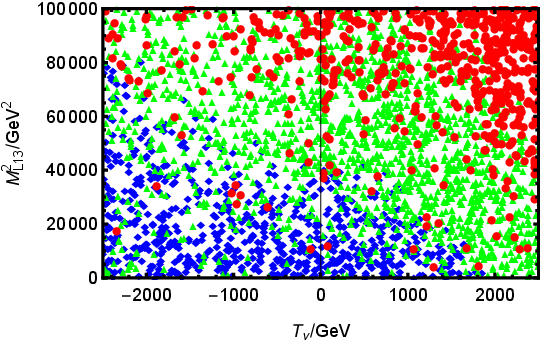}
\caption{Under the constraint of Br($\tau\rightarrow 3e$) process, a reasonable parameter space is selected for random scattering, and the marking of the scattering points represents: $\textcolor{blue}{\blacklozenge}~(0<Br(\tau\rightarrow 3e)<1\times 10^{-11}),~\textcolor{green}{\blacktriangle}~(1\times 10^{-11}\leq Br(\tau\rightarrow 3e)<8\times 10^{-11}),~ \textcolor{red}{\bullet}~(8\times 10^{-11}\leq Br(\tau\rightarrow 3e)<2.7\times 10^{-8})$.}{\label {tao3e1}}
\end{figure}

\begin{figure}[ht]
\setlength{\unitlength}{5mm}
\centering
\includegraphics[width=2.9in]{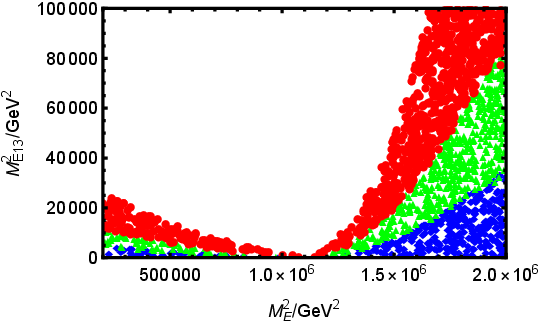}
\caption{Under the constraint of Br($\tau\rightarrow 3e$) process, a reasonable parameter space is selected for random scattering, and the marking of the scattering points represents: $\textcolor{blue}{\blacklozenge}~(0<Br(\tau\rightarrow 3e)<1.3\times 10^{-11}),~\textcolor{green}{\blacktriangle}~(1.3\times 10^{-11}\leq Br(\tau\rightarrow 3e)<8\times 10^{-11}),~ \textcolor{red}{\bullet}~(8\times 10^{-11}\leq Br(\tau\rightarrow 3e)<2.7\times 10^{-8})$.}{\label {tao3e2}}
\end{figure}

In the same way, we analyze the LFV process $\tau\rightarrow 3e$, whose experimental upper bound is Br$(\tau\rightarrow 3e)<2.7\times10^{-8}$. In order to get numerical results for $\tau\rightarrow 3e$, we use $\tan\beta=25,~g_B=0.3,~g_{YB}=0.1,~\lambda=0.4,~v_S=4~{\rm TeV},~M_2=1.2~{\rm TeV}, ~M_{L}^2=1.6\times10^{5}~{\rm GeV^2}$ and carry out two random scattering points to obtain the influence of $M_{E}^2, ~M_{E13}^2,~T_\nu,~M_{L13}^2$ on Br$(\tau\rightarrow 3e)<2.7\times10^{-8}$ in Fig.\ref{tao3e1} and Fig.\ref{tao3e2}.

The parameters are randomly scanned for the first time. The ranges of parameters are shown in Table \ref{biao4}, and the branching ratio of $\tau\rightarrow 3e$ process is expressed as: $\textcolor{blue}{\blacklozenge}~(0<Br(\tau\rightarrow \mu\gamma)<1\times 10^{-11}),~\textcolor{green}{\blacktriangle}~(1\times 10^{-11}\leq Br(\tau\rightarrow \mu\gamma)<8\times 10^{-11}),~ \textcolor{red}{\bullet}~(8\times 10^{-11}\leq Br(\tau\rightarrow \mu\gamma)<2.7\times 10^{-8})$.

\begin{table*}
\caption{Scanning parameters for Fig.{\ref {tao3e1}}}\label{biao4}
\begin{tabular}{|c|c|c|}
\hline
Parameters&Min&Max\\
\hline
$\hspace{1.5cm}M^2_{L}/\rm GeV^2\hspace{1.5cm}$ &$\hspace{1.5cm}2\times10^{5}\hspace{1.5cm}$& $\hspace{1.5cm}2\times10^{6}\hspace{1.5cm}$\\
\hline
$\hspace{1.5cm}M^2_{L13}/\rm GeV^2\hspace{1.5cm}$ &$\hspace{1.5cm}0\hspace{1.5cm}$& $\hspace{1.5cm}10^5\hspace{1.5cm}$\\
\hline
$\hspace{1.5cm}T_{\nu}/\rm GeV\hspace{1.5cm}$ &$\hspace{1.5cm}-2500\hspace{1.5cm}$ &$\hspace{1.5cm}2500\hspace{1.5cm}$\\
\hline
$T_{e 13}$/GeV & $\hspace{1.5cm}-500\hspace{1.5cm}$ &$\hspace{1.5cm}500\hspace{1.5cm}$\\
\hline
$\tan\beta$ &$\hspace{1.5cm}5\hspace{1.5cm}$ &$\hspace{1.5cm}50\hspace{1.5cm}$\\
\hline
\end{tabular}
\end{table*}

Fig.\ref{tao3e1} illustrates the distributional properties of the branching ratio of $\tau\rightarrow 3e$ in the ($T_\nu,~M_{L13}^2$) parameter space. The analysis indicates that when $M_{L13}^2$  is held constant, Br$(\tau\rightarrow 3e)$ significantly increases with the rise of $T_\nu$, demonstrating that positive  $T_\nu$  has a pronounced effect on enhancing flavor violation effects. When $T_\nu$ is fixed, Br$(\tau\rightarrow 3e)$ markedly grows with the increase of $M_{L13}^2$, indicating that the non-diagonal mass term  $M_{L13}^2$  is one of the dominant factors in flavor violation effects.

Then we continue to randomly scatter points, and the parameter ranges are shown in Table \ref{biao5}. $\textcolor{blue}{\blacklozenge}$, $\textcolor{green}{\blacktriangle}$ and $\textcolor{red}{\bullet}$ indicate the range $0<Br(\tau\rightarrow 3e)<1.3\times 10^{-11}$, $1.3\times 10^{-11}\leq Br(\tau\rightarrow 3e)<8\times 10^{-11}$, $8\times 10^{-11}\leq Br(\tau\rightarrow 3e)<2.7\times 10^{-8}$.

\begin{table*}
\caption{Scanning parameters for Fig.{\ref {tao3e2}}}\label{biao5}
\begin{tabular}{|c|c|c|}
\hline
Parameters&Min&Max\\
\hline
$\hspace{1.5cm}M^2_{L}/\rm GeV^2\hspace{1.5cm}$ &$\hspace{1.5cm}2\times10^{5}\hspace{1.5cm}$& $\hspace{1.5cm}2\times10^{6}\hspace{1.5cm}$\\
\hline
$\hspace{1.5cm}M^2_{E}/\rm GeV^2\hspace{1.5cm}$ &$\hspace{1.5cm}2\times10^{5}\hspace{1.5cm}$& $\hspace{1.5cm}2\times10^{6}\hspace{1.5cm}$\\
\hline
$\hspace{1.5cm}M^2_{\nu}/\rm GeV^2\hspace{1.5cm}$ &$\hspace{1.5cm}1\times10^{5}\hspace{1.5cm}$& $\hspace{1.5cm}2\times10^{6}\hspace{1.5cm}$\\
\hline
$\hspace{1.5cm}M^2_{E13}/\rm GeV^2\hspace{1.5cm}$ &$\hspace{1.5cm}0\hspace{1.5cm}$& $\hspace{1.5cm}10^5\hspace{1.5cm}$\\
\hline
$\hspace{1.5cm}T_{e}/\rm GeV\hspace{1.5cm}$ &$\hspace{1.5cm}-2500\hspace{1.5cm}$ &$\hspace{1.5cm}2500\hspace{1.5cm}$\\
\hline
$\hspace{1.5cm}T_{\nu}/\rm GeV\hspace{1.5cm}$ &$\hspace{1.5cm}-2500\hspace{1.5cm}$ &$\hspace{1.5cm}2500\hspace{1.5cm}$\\
\hline
\end{tabular}
\end{table*}

The distribution of Br$(\tau\rightarrow 3e)$ on the ($M_{E}^2$, $M_{E13}^2$) plane shows an asymmetric  "U" shape with some blank areas in Fig.\ref{tao3e2}. The asymmetry of the "U" shaped distribution in Fig.\ref{tao3e2} is attributed to the predominance of the diagonal term $M_{E}^2$: when $M_{E}^2$ is small, the slepton mass is relatively light, and the influence of the non-diagonal term on the flavor violation effect is more pronounced, making the red region more likely to appear at low $M_{E}^2$ with a large $M_{E13}^2$. Conversely, when $M_{E}^2$ is larger, the diagonal term dominates the slepton mass eigenvalues, and the contribution of the non-diagonal term is relatively suppressed, necessitating a larger $M_{E13}^2$ for the red region to emerge. It is evident that the action of $M_{E}^2$ on Br$(\tau\rightarrow 3e)$ changes differently on both sides of $M_{E}^2 \approx 1.1\times 10^{6}~{\rm GeV^2}$. For smaller values of $M_{E}^2$  $(2\times 10^{5}~{\rm GeV}^2\leq M_{E}^2<1.1\times 10^{6}~{\rm GeV}^2)$, the mass of slepton is lighter, which leads to stronger flavor violation effect. As $M_{E}^2$ increases, the interactions among supersymmetric particles, such as Chargino ($\chi^{\pm}$) with sneutrino ($\tilde{\nu}$) and neutralino ($\chi^{0}$) with slepton ($\tilde{L}$), may become more significant, resulting in the branching ratio to enhance with $M_{E}^2$. This trend is contrary to the usual expectation, likely because the increase arise from amplified coupling strengths between these particles, intensifying flavor violation effect. In contrast, at larger $M_{E}^2$ $(1.1\times 10^{6}~{\rm GeV}^2\leq M_{E}^2<2\times 10^{6}~{\rm GeV}^2)$, slepton mass grows sufficiently heavy to suppress flavor violation, causing Br($\tau\rightarrow 3e$) to decrease with further increases in $M_{E}^2$, aligning with standard predictions. The peak observed near $M_{E}^2 \approx 1.1\times 10^{6}~{\rm GeV^2}$ corresponds to a maximal branching ratio. Here, slepton mass is elevated but not yet heavy enough to fully suppress flavor violation. At this critical point, interactions between $\chi^{0}$$\tilde{\nu}$ and $\chi^{0}$$\tilde{L}$ reach their strongest effective couping, optimizing flavor violation contributions to $\tau\rightarrow 3e$. $M_{E13}^2$ is the non-diagonal term in the slepton mass matrix and governs the mixing between different generations of slepton. Larger $M_{E13}^2$ enhances this mixing, typically amplifying flavor violation effects and thereby increasing Br$(\tau\rightarrow 3e)$. In conclusion, when the vertical axis is a constant value, considering the effect of the horizontal axis on Br$(\tau\rightarrow 3e)$ that the leftward rise in the branching ratio is likely driven by $\chi^{0}$$\tilde{\nu}$ and $\chi^{0}$$\tilde{L}$ interactions, while the subsequent decline at higher $M_{E}^2$ reflects the weakening of $\chi^{0}$$\tilde{L}$ couplings due to heavier slepton. The emergence of white region in Fig.\ref{tao3e2} may be attributed to several factors: the particles masses may exceed the detectable ranges of the LHC; the flavor violation effect may be too significant, surpassing the upper limits of experimental measurements.

\subsection{$\mu\rightarrow e+q\bar q$}

In this subsection, we discuss numerical results for $\mu\rightarrow e+q\bar q$ and consider constraints from the LFV processes $l_j\rightarrow l_i\gamma$ and $l_j\rightarrow 3l_i$. In this work, we use the parameters $T_\nu=1~{\rm TeV}, T_e=1~{\rm TeV}, v_S=4~{\rm TeV},  M_{L}^2=1.6\times10^{5}~{\rm GeV^2}$.

\begin{table*}
\caption{Scanning parameters for Fig.{\ref {mueqq1}} and Fig.{\ref {mueqq3}}}\label{biao6}
\begin{tabular}{|c|c|c|}
\hline
Parameters&Min&Max\\
\hline
$\hspace{1.5cm}M^2_{L}/\rm GeV^2\hspace{1.5cm}$ &$\hspace{1.5cm}2\times10^{5}\hspace{1.5cm}$& $\hspace{1.5cm}2\times10^{6}\hspace{1.5cm}$\\
\hline
$\hspace{1.5cm}M^2_{E}/\rm GeV^2\hspace{1.5cm}$ &$\hspace{1.5cm}2\times10^{5}\hspace{1.5cm}$& $\hspace{1.5cm}2\times10^{6}\hspace{1.5cm}$\\
\hline
$\hspace{1.5cm}M^2_{\nu}/\rm GeV^2\hspace{1.5cm}$ &$\hspace{1.5cm}1\times10^{5}\hspace{1.5cm}$& $\hspace{1.5cm}2\times10^{6}\hspace{1.5cm}$\\
\hline
$\hspace{1.5cm}M^2_{E12}/\rm GeV^2\hspace{1.5cm}$ &$\hspace{1.5cm}0\hspace{1.5cm}$& $\hspace{1.5cm}100\hspace{1.5cm}$\\
\hline
$\hspace{1.5cm}M_2/\rm GeV\hspace{1.5cm}$ &$\hspace{1.5cm}900\hspace{1.5cm}$ &$\hspace{1.5cm}3000\hspace{1.5cm}$\\
\hline
$\hspace{1.5cm}v_S/\rm GeV\hspace{1.5cm}$ &$\hspace{1.5cm}2000\hspace{1.5cm}$ &$\hspace{1.5cm}7000\hspace{1.5cm}$\\
\hline
$\hspace{1.5cm}T_{e12}/\rm GeV\hspace{1.5cm}$ &$\hspace{1.5cm}0\hspace{1.5cm}$ &$\hspace{1.5cm}10\hspace{1.5cm}$\\
\hline
$\hspace{1.5cm}\tan\beta\hspace{1.5cm}$ &$\hspace{1.5cm}5\hspace{1.5cm}$
&$\hspace{1.5cm}50\hspace{1.5cm}$\\
\hline
$\hspace{1.5cm}g_B\hspace{1.5cm}$ &$\hspace{1.5cm}0.3\hspace{1.5cm}$
&$\hspace{1.5cm}0.6\hspace{1.5cm}$\\
\hline
$\hspace{1.5cm}g_{YB}\hspace{1.5cm}$ &$\hspace{1.5cm}0.01\hspace{1.5cm}$
&$\hspace{1.5cm}0.2\hspace{1.5cm}$\\
\hline
$\hspace{1.5cm}\lambda\hspace{1.5cm}$ &$\hspace{1.5cm}0.1\hspace{1.5cm}$
&$\hspace{1.5cm}0.4\hspace{1.5cm}$\\
\hline
\end{tabular}
\end{table*}

\begin{figure}[ht]
\setlength{\unitlength}{5mm}
\centering
\includegraphics[width=2.9in]{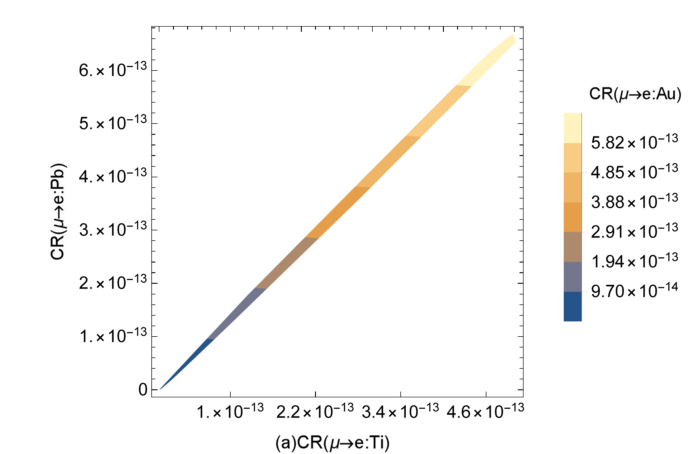}
\setlength{\unitlength}{5mm}
\centering
\includegraphics[width=2.9in]{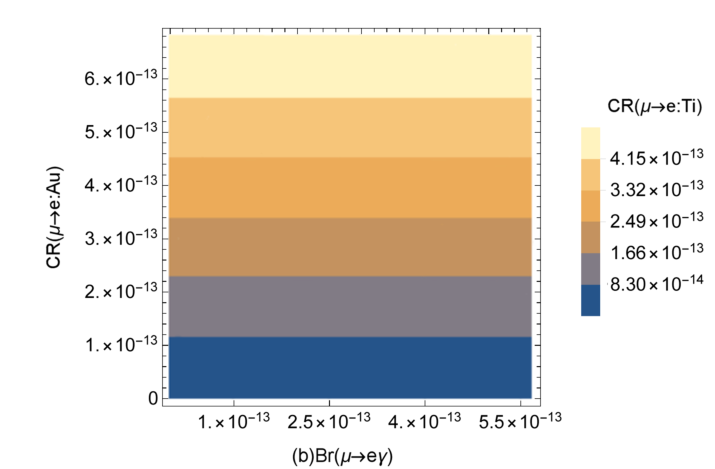}
\setlength{\unitlength}{5mm}
\centering
\includegraphics[width=2.9in]{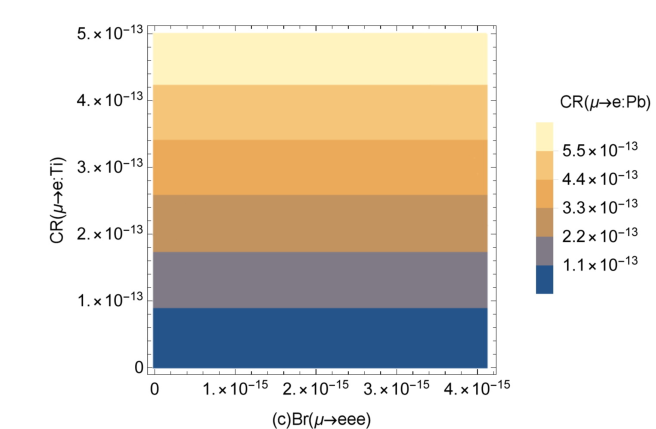}
\caption{(a) shows the relationship of CR$(\mu\rightarrow e$:Ti) and CR$(\mu\rightarrow e$:Pb) on CR$(\mu\rightarrow e$:Au). (b) shows the relationship of Br$(\mu\rightarrow e\gamma)$ and CR$(\mu\rightarrow e$:Au) on CR$(\mu\rightarrow e$:Ti). (c)shows the relationship of Br$(\mu\rightarrow eee)$ and CR$(\mu\rightarrow e$:Ti) on CR$(\mu\rightarrow e$:Pb).}{\label {mueqq1}}
\end{figure}

In Fig.\ref{mueqq1}, we set $\tan\beta=25,~\lambda=0.4,~g_B=0.3,~g_{YB}=0.1,~M_2=1.2~{\rm TeV},~M_{E}^2=1.5{\rm TeV^2}$ and scan some parameters in Table \ref{biao6}. Fig.\ref{mueqq1}(a) demonstrates a strong correlation among CR$(\mu\rightarrow e$:Ti), CR$(\mu\rightarrow e$:Pb) and  CR$(\mu\rightarrow e$:Au). The conversion rates of the three are similar by orders of magnitude and have the same growth trend. Therefore, in order to simplify the analysis, it is possible to study only the behavior of CR$(\mu\rightarrow e$:Au) without affecting the understanding of the overall physical mechanism. The values of CR$(\mu\rightarrow e$:Au) are calculated by satisfying the two experimental constraints Br$(\mu\rightarrow e\gamma)$ and Br$(\mu\rightarrow eee)$. Fig.\ref{mueqq1}(b) and Fig.\ref{mueqq1}(c) remains stable under these constraints, their values are not very closely related to Br$(\mu\rightarrow e\gamma)$ and Br$(\mu\rightarrow eee)$. This result may reflect the relative independence of different decay processes in a particular parameter space. Based on the above analysis, in order to save space, we only use CR$(\mu\rightarrow e$:Au) for the subsequent discussion when CR$(\mu\rightarrow e$:Ti)$<4.3\times10^{-12}$, CR$(\mu\rightarrow e$:Pb)$<4.6\times10^{-11}$ are satisfied, without affecting the universality of the conclusions.

\begin{figure}[ht]
\setlength{\unitlength}{5mm}
\centering
\includegraphics[width=2.9in]{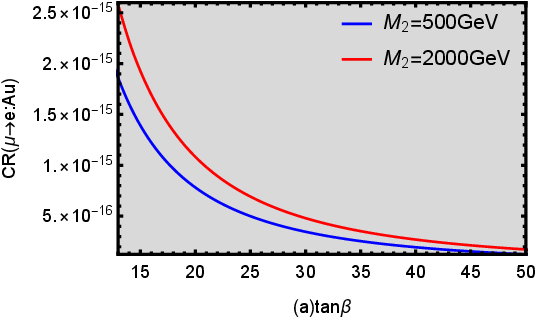}
\setlength{\unitlength}{5mm}
\centering
\includegraphics[width=2.9in]{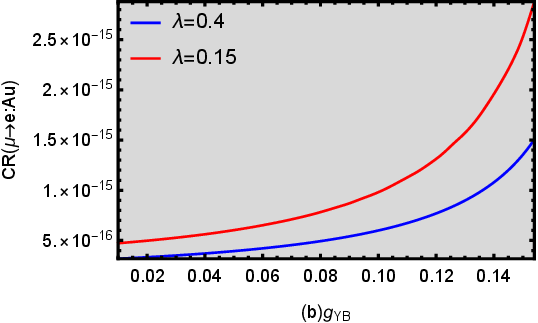}
\setlength{\unitlength}{5mm}
\centering
\includegraphics[width=2.9in]{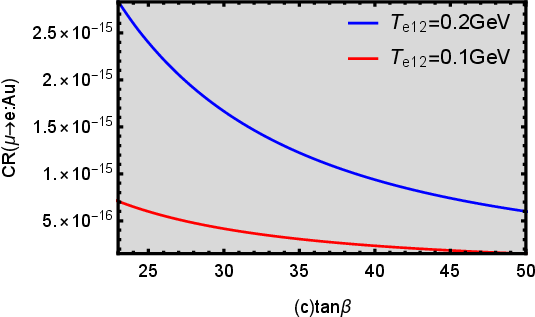}
\caption{CR$(\mu\rightarrow e$:Au) schematic diagrams affected by different parameters, the gray area satisfies the experimental upper limit. Fig.\ref{mueqq2}(a) shows the relationship between $\tan\beta$ and CR$(\mu\rightarrow e$:Au) with the red line representing $M_2=2000~{\rm GeV}$ and the blue line representing  $M_2=500~{\rm GeV}$. Fig.\ref{mueqq2}(b) shows the relationship between $g_{YB}$ and CR$(\mu\rightarrow e$:Au) with the red line representing $\lambda=0.15$ and the blue line representing $\lambda=0.4$. Fig.\ref{mu3e}(c) shows the relationship between $\tan\beta$ and CR$(\mu\rightarrow e$:Au) with the red line representing $T_{e12}=0.1~{\rm GeV}$ and the blue line representing $T_{e12}=0.2~{\rm GeV}$.}{\label {mueqq2}}
\end{figure}

With the parameters $\lambda=0.4,~g_B=0.3,~g_{YB}=0.1,~T_{e12}=0.1~{\rm GeV},~M_{E}^2=1.5~{\rm TeV^2}$, Fig.\ref{mueqq2}(a) shows CR$(\mu\rightarrow e$:Au) as a function of $\tan\beta$. It can be seen that CR$(\mu\rightarrow e$:Au) decreases as $\tan\beta$ increases. In addition, $M_2=2000~{\rm GeV}$ makes CR$(\mu\rightarrow e$:Au) significantly higher than the case of $M_2=500~{\rm GeV}$. Supposing $\tan\beta=25, ~g_B=0.3,~T_{e12}=0.1~{\rm GeV},~M_2=1.2~{\rm TeV},~M_{E}^2=1.5~{\rm TeV^2}$, Fig.\ref{mueqq2}(b) examines the trend of CR$(\mu\rightarrow e$:Au) as $g_{YB}$ varies, considering two different values of $\lambda$: $\lambda=0.4$ (blue line) and $\lambda=0.15$ (red line). As can be seen from Fig.\ref{mueqq2}(b), CR$(\mu\rightarrow e$:Au) rises monotonically with increasing $g_{YB}$ and smaller $\lambda$ corresponds to larger conversion rates. Setting $\lambda=0.4,~g_B=0.3,~g_{YB}=0.1,~M_2=1.2~{\rm TeV},~M_{E}^2=1.5~{\rm TeV^2}$, we plot CR$(\mu\rightarrow e$:Au) versus $\tan\beta$ in Fig.\ref{mueqq2}(c), the red line corresponds to $T_{e 12}=0.1~{\rm GeV}$ and the blue line corresponds to $T_{e 12}=0.2~{\rm GeV}$. We can clearly see that the two lines decrease with the increasing $\tan\beta$, and the conversion rate of $T_{e12}=0.2~{\rm GeV}$ (blue line) is always higher than $T_{e12}=0.1~{\rm GeV}$ (red line). In conclusion, Fig.\ref{mueqq2}(a) and Fig.\ref{mueqq2}(c) show that an increase in $\tan\beta$ leads to a decrease in CR$(\mu\rightarrow e$:Au). Fig.\ref{mueqq2}(b) shows that $g_{YB}$, as an important additional parameter, can enhance the LFV process, resulting in a significant increase in conversion rate. In addition, different values of the parameters $M_2$, $\lambda$ and $T_{e12}$ also affect the magnitude of CR$(\mu\rightarrow e$:Au), suggesting that these parameters play a key role in LFV.

\begin{figure}[ht]
\setlength{\unitlength}{5mm}
\centering
\includegraphics[width=2.9in]{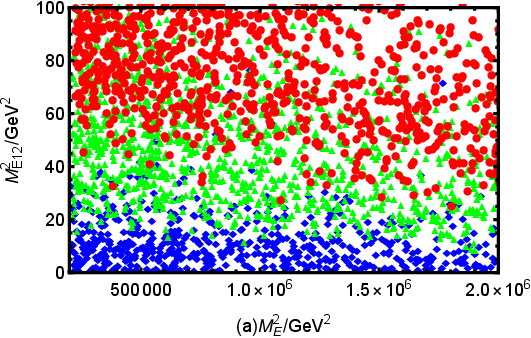}
\setlength{\unitlength}{5mm}
\centering
\includegraphics[width=2.7in]{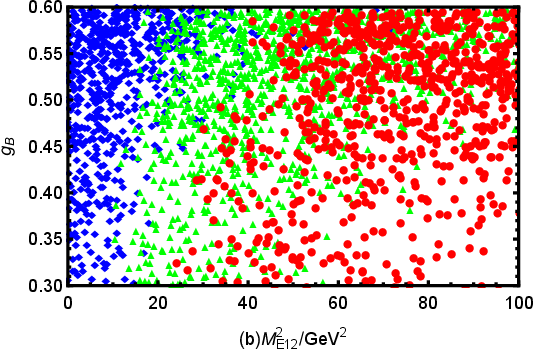}
\caption{Under the constraint of CR$(\mu\rightarrow e$:Au) process, a reasonable parameter space is selected for random scattering, and the marking of the scattering points represents: $\textcolor{blue}{\blacklozenge}~(0<$CR$(\mu\rightarrow e$:Au)$<3\times 10^{-14})$,$~\textcolor{green}{\blacktriangle}~(3\times 10^{-14}\leq$ CR$(\mu\rightarrow e$:Au)$<2.8\times 10^{-13})$,$~ \textcolor{red}{\bullet}~(2.8\times 10^{-13}\leq $CR$(\mu\rightarrow e$:Au)$<7\times 10^{-13})$.}{\label {mueqq3}}
\end{figure}

Next, we randomly scan the parameters. Based on $\tan\beta=25,~\lambda=0.4,~g_{YB}=0.1,~M_2=1.2~{\rm TeV}$, Fig.\ref{mueqq3} is obtained from the parameters shown in Table \ref{biao6}. We use $\textcolor{blue}{\blacklozenge}~(0<$CR$(\mu\rightarrow e$:Au)$<3\times 10^{-14})$,$~\textcolor{green}{\blacktriangle}~(3\times 10^{-14}\leq$ CR$(\mu\rightarrow e$:Au)$<2.8\times 10^{-13})$,$~ \textcolor{red}{\bullet}~(2.8\times 10^{-13}\leq $CR$(\mu\rightarrow e$:Au)$<7\times 10^{-13})$ to represent the results in different parameter spaces for the process of $\mu\rightarrow e+q\bar q$. The relationship between $M_{E}^2$ and $M_{E12}^2$ is shown in Fig.\ref{mueqq3}(a). It is evident from the distribution of data points that CR$(\mu\rightarrow e$:Au) rises significantly with the increase of $M_{E12}^2$, while the dependence on $M_{E}^2$ is relatively weak. Specifically, $\textcolor{blue}{\blacklozenge}$ are mainly in $0~{\rm GeV^2}<M_{E12}^2<20~{\rm GeV^2}$, $\textcolor{green}{\blacktriangle}$ are mainly in $20~{\rm GeV^2}<M_{E12}^2<60~{\rm GeV^2}$ and $ \textcolor{red}{\bullet}$ are mainly in $60~{\rm GeV^2}<M_{E12}^2<100~{\rm GeV^2}$. This trend is consistent with the expectation from LFV theory that the non-diagonal term $M_{E12}^2$ directly controls the flavor mixing strength of the slepton mass matrix, thereby enhancing the amplitude of the jump of the LFV process leading to a larger CR$(\mu\rightarrow e$:Au). In contrast, $M_{E}^2$ mainly affects the slepton mass scale, and its effect on CR$(\mu\rightarrow e$:Au) is small within the given parameter ranges, so the data points do not show a significant trend in the $M_{E}^2$ direction. Fig.\ref{mueqq3}(b) shows the distribution of CR$(\mu\rightarrow e$:Au) for different $M_{E12}^2$ and $g_B$ values. One can see that CR$(\mu\rightarrow e$:Au) primarily rises with $M_{E12}^2$, bringing about the same effect as in Fig.\ref{mueqq3}(a). The effect of $g_B$ on the conversion rate is also present to some extent, although its effect is not as significant as that of $M_{E12}^2$. In the larger $g_B$ region, the density of data points is higher, while in the smaller $g_B$ region, the data points are more sparse.

\section{discussion and conclusion}

In the SM, the LFV decays $l_j\rightarrow l_i\gamma$ and $l_j \rightarrow 3l_i$ have extremely low predicted branching ratios. For example, the branching ratio Br$(\mu \rightarrow e\gamma) \sim 10^{-55}$ is much lower than the current experimental upper limit of $10^{-13}$, rendering these decays virtually unobservable in the SM. Similarly, the $\mu\rightarrow e+ q\bar q$ conversion rate is predicted to be negligibly small. Therefore, if signals of these LFV processes are experimentally detected, they would necessarily indicate NP beyond the SM. Within the framework of the N-B-LSSM, we analyze the contributions of the newly introduced particles and couplings to LFV processes by calculating the corresponding Feynman diagrams and performing an extensive parameter space scan. Compared with the MSSM, the N-B-LSSM introduces additional superfields, including right-handed neutrinos and three Higgs superfields, which not only help resolve certain issues in the MSSM but also provide extra sources for LFV, thereby significantly enhancing the LFV signals.

The experimental limits on the branching ratios for $\mu \rightarrow e\gamma$ and $\mu \rightarrow 3e$, as well as on the conversion rate for $\mu\rightarrow e+ q\bar q$ are extremely stringent, which strongly constrains the theoretical parameter space. In contrast, the experimental upper limits for $\tau \rightarrow e\gamma$, $\tau \rightarrow \mu\gamma$, $\tau \rightarrow 3e$ and $\tau \rightarrow 3\mu$ are around $10^{-8}$ orders of magnitude, imposing relatively looser constraints. Our results indicate that parameters such as $\tan\beta$,~$g_B$,~$g_{YB}$,~$v_S$,~$\lambda$, ~$T_{e}$,~$T_{\nu}$,~$T_{e ij}$,~$T_{\nu ij}$,~$M_2$,~$M_{L}^2$,~$M_{E}^2$,~$M_{Lij}^2$,~$M_{Eij}^2$ and $M_{\nu ij}^2$~$(i,j=1,2,3,~i\neq j)$ have varying degrees of influence on LFV processes, with $M_{Lij}^2$,~ $M_{Eij}^2$,~$M_{\nu ij}^2$,~$T_{eij}$,~$T_{\nu ij}$,~$g_B$,~$g_{YB}$ and $\tan\beta$ being particularly sensitive. In conclusion, through the analysis of lepton flavour mixing parameters, we find that the non-diagonal elements which correspond to the generations of the initial lepton and final lepton are main sensitive parameters and LFV sources. Most of the parameters are able to break the experimental upper limit, providing new ideas for the search of NP.

\begin{acknowledgments}
This work is supported by National Natural Science Foundation of China (NNSFC)
(No.12075074), Natural Science Foundation of Hebei Province
(A2023201040, A2022201022, A2022201017, A2023201041), Natural Science Foundation of Hebei Education Department (QN2022173), Post-graduate's Innovation Fund Project of Hebei University (HBU2024SS042), the Project of the China Scholarship Council (CSC) No. 202408130113.
\end{acknowledgments}

\appendix
\section{Mass matrix and coupling  in N-B-LSSM}\label{A1}

The mass matrix for neutralino $(\lambda_{\tilde{B}}, \tilde{W}^0, \tilde{H}_d^0, \tilde{H}_u^0, \tilde{B'}, \tilde{\chi_1}, \tilde{\chi_2}, S)$ reads:

\begin{equation}
m_{{\chi}^0}= \left(
\begin{array}{cccccccc}
M_1 &0 &-\frac{1}{2}g_1 v_d &\frac{1}{2} g_1 v_u &{M}_{B B'} & 0  & 0  &0\\
0 &M_2 &\frac{1}{2} g_2 v_d  &-\frac{1}{2} g_2 v_u  &0 &0 &0 &0\\
-\frac{1}{2}g_1 v_d &\frac{1}{2} g_2 v_d  &0
&- \frac{1}{\sqrt{2}} {\lambda} v_S&-\frac{1}{2} g_{YB} v_d &0 &0 & - \frac{1}{\sqrt{2}} {\lambda} v_u\\
\frac{1}{2}g_1 v_u &-\frac{1}{2} g_2 v_u  &- \frac{1}{\sqrt{2}} {\lambda} v_S &0 &\frac{1}{2} g_{YB} v_u  &0 &0 &- \frac{1}{\sqrt{2}} {\lambda} v_d\\
{M}_{B B'} &0 &-\frac{1}{2} g_{YB} v_{d}  &\frac{1}{2} g_{YB} v_{u} &{M}_{BL} &- g_{B} v_{\eta}  &g_{B} v_{\bar{\eta}}  &0\\
0  &0 &0 &0 &- g_{B} v_{\eta}  &0 &-\frac{1}{\sqrt{2}} {\lambda}_{2} v_S  &-\frac{1}{\sqrt{2}} {\lambda}_{2} v_{\bar{\eta}} \\
0 &0 &0 &0 &g_{B} v_{\bar{\eta}}  &-\frac{1}{\sqrt{2}} {\lambda}_{2} v_S  &0 &-\frac{1}{\sqrt{2}} {\lambda}_{2} v_{\eta} \\
0 &0 & - \frac{1}{\sqrt{2}} {\lambda} v_u &- \frac{1}{\sqrt{2}}{\lambda} v_d &0 &-\frac{1}{\sqrt{2}} {\lambda}_{2} v_{\bar{\eta}}
 &-\frac{1}{\sqrt{2}} {\lambda}_{2} v_{\eta}  &\sqrt{2}\kappa v_S\end{array}
\right).
 \end{equation}

This matrix is diagonalized by the rotation matrix $N$,
\begin{equation}
N^{*}m_{{\chi}^0} N^{\dagger} =m^{dia}_{{\chi}^0}.
\end{equation}

In the basis $(\tilde{W}^-,\tilde{H}_d^-)$ and $(\tilde{W}^+,\tilde{H}_u^+)$, the definition of the mass matrix for chargino is given by:
\begin{eqnarray}
m_{\tilde{\chi}^-} = \left(
\begin{array}{cc}
M_2&\frac{1}{\sqrt{2}}g_2v_u\\
\frac{1}{\sqrt{2}}g_2v_d&\frac{1}{\sqrt{2}}\lambda v_S\end{array}
\right).
\end{eqnarray}

This matrix is diagonalized by U and V:
\begin{eqnarray}
U^{*} m_{\tilde{\chi}^-} V^{\dagger}= m_{\tilde{\chi}^-}^{dia}.
\end{eqnarray}

The mass matrix for slepton in the basis $(\tilde{e_L},\tilde{e_R}),(\tilde{e_L}^*,\tilde{e_R}^*)$ is:
\begin{eqnarray}
m^2_{\tilde{e}} = \left(
\begin{array}{cc}
m_{\tilde{e}_L\tilde{e}_L^*} &\frac{1}{\sqrt{2}}  v_d T_{e}^{\dagger}  -\frac{1}{2} {\lambda} v_S v_u Y_{e}^{\dagger} \\
\frac{1}{\sqrt{2}} v_d T_{e} - \frac{1}{2} v_S v_u Y_{e} {\lambda}^*  &m_{\tilde{e}_R\tilde{e}_R^*}\end{array}
\right),
\end{eqnarray}
\begin{eqnarray}
&&m_{\tilde{e}_L\tilde{e}_L^*} =\frac{1}{8} \Big((g_{1}^{2} + g_{Y B}^{2}
+ g_{Y B} g_{B} -g_2^2)(v_{d}^{2}- v_{u}^{2})+ 2(g_{B}^2+ g_{Y B} g_{B})( v_{\eta}^{2}- v_{\bar{\eta}}^{2}
)
\Big)\nonumber\\&&\hspace{1.1cm}+\frac{1}{2} v_{d}^{2}Y_{e}^{\dagger}Y_{e}+m_{\tilde{L}}^2 ,\nonumber\\&&
m_{\tilde{e}_R\tilde{e}_R^*} = \frac{1}{8}\Big((2g_{1}^{2}+2g_{Y B}^{2}+g_{Y B} g_{B})
(v_{u}^{2}-v_{d}^{2})+2(2g_{Y B} g_{B}+g_{B}^{2})(v_{\bar{\eta}}^{2}-v_{\eta}^{2})
\Big)\nonumber\\&&\hspace{1.1cm}+\frac{1}{2} v_{d}^{2}Y_{e}Y_{e}^{\dagger}+m_{\tilde{E}}^2.
\end{eqnarray}

The unitary matrix $Z^E$ is used to rotate slepton mass squared matrix to mass eigenstates:
\begin{eqnarray}
Z^E m^2_{\tilde{e}} Z^{E,\dagger} = m^{dia}_{2,\tilde{e}}.
\end{eqnarray}

The mass squared matrix for CP-even sneutrino $({\phi}_{l}, {\phi}_{r})$ reads:
\begin{eqnarray}
m^2_{\tilde{\nu}^R} = \left(
\begin{array}{cc}
m_{{\phi}_{l}{\phi}_{l}} &m^T_{{\phi}_{r}{\phi}_{l}}\\
m_{{\phi}_{l}{\phi}_{r}} &m_{{\phi}_{r}{\phi}_{r}}\end{array}
\right),
\end{eqnarray}
\begin{eqnarray}
&&m_{{\phi}_{l}{\phi}_{l}}= \frac{1}{8} \Big((g_{1}^{2} + g_{Y B}^{2} + g_{2}^{2}+  g_{Y B} g_{B})( v_{d}^{2}- v_{u}^{2})
+  2(g_{Y B} g_{B}+g_{B}^2)(v_{\eta}^{2}-v_{\bar{\eta}}^{2})\Big)
\nonumber\\&&\hspace{1.1cm}+\frac{1}{2} v_{u}^{2}{Y_{\nu}^{T}  {Y_\nu}^*}  + m_{\tilde{L}}^2,
\\&&m_{{\phi}_{l}{\phi}_{r}} =-\frac{1}{2}v_d v_S Y_\nu \lambda^* + v_u v_{\eta} {Y_X  {Y_\nu}^*}+\frac{1}{\sqrt{2}}v_u T_\nu,\\&&
m_{{\phi}_{r}{\phi}_{r}}= \frac{1}{8} \Big(g_{Y B} g_{B}(v_{u}^{2}- v_{d}^{2})
+2g_{B}^{2}(v_{\bar{\eta}}^{2}-v_{\eta}^{2})\Big)-v_S v_{\bar{\eta}}Y_X {{\lambda}_{2}}^*+m_{\tilde{\nu}}^2+\frac{1}{2} v_{u}^{2}Y_\nu {Y_\nu}^\dagger\nonumber \\&&\hspace{1.1cm}
+ v_{\eta} (2 v_{\eta}Y_X Y_X^* + \sqrt{2} T_X).
\end{eqnarray}

To obtain the mass of CP-even sneutrino, we diagonalize the matrix $m_{\tilde\nu^R}^2$ using the rotation matrix $Z^R$:

\begin{eqnarray}
Z^R m^2_{\tilde{\nu}^R} Z^{R,\dagger} = m^{dia}_{2,\tilde{\nu}^R}.
\end{eqnarray}

The mass squared matrix for CP-odd sneutrino $({\sigma}_{l}, {\sigma}_{r})$ is also derived here:

\begin{eqnarray}
m^2_{\tilde{\nu}^I} = \left(
\begin{array}{cc}
m_{{\sigma}_{l}{\sigma}_{l}} &m^T_{{\sigma}_{r}{\sigma}_{l}}\\
m_{{\sigma}_{l}{\sigma}_{r}} &m_{{\sigma}_{r}{\sigma}_{r}}\end{array}
\right),
\end{eqnarray}
\begin{eqnarray}
&&m_{{\sigma}_{l}{\sigma}_{l}}= \frac{1}{8} \Big((g_{1}^{2} + g_{Y B}^{2} + g_{2}^{2}+  g_{Y B} g_{B})( v_{d}^{2}- v_{u}^{2})+2(g_B^2+g_{Y B} g_{B})(v_{\eta}^{2}-v_{\bar{\eta}}^{2})\Big)
\nonumber\\&&\hspace{1.1cm}+\frac{1}{2} v_{u}^{2}{Y_{\nu}^{T}  Y_\nu^*}  + m_{\tilde{L}}^2,
\\&&m_{{\sigma}_{l}{\sigma}_{r}} = -\frac{1}{2} v_d v_S Y_\nu \lambda^*-v_u v_{\eta}{Y_X Y_\nu^*}+\frac{1}{\sqrt{2}} v_u T_\nu,\\&&
m_{{\sigma}_{r}{\sigma}_{r}}= \frac{1}{8} \Big(g_{Y B} g_{B}(v_{u}^{2}- v_{d}^{2})+2g_{B}^{2}(v_{\bar{\eta}}^{2}-v_{\eta}^{2})\Big)+v_S v_{\bar{\eta}} Y_X {\lambda}_{2}^*+m_{\tilde{\nu}}^2 + \frac{1}{2} v_{u}^{2}Y_\nu {Y_\nu}^\dagger \nonumber \\&&\hspace{1.1cm}
+ v_{\eta} (2 v_{\eta}Y_X  Y_X^*  - \sqrt{2} T_X).
\end{eqnarray}

This matrix is diagonalized by $Z^I$:

\begin{eqnarray}
Z^I m^2_{\tilde{\nu}^I} Z^{I,\dagger} = m^{dia}_{2,\tilde{\nu}^I}.
\end{eqnarray}

In the basis
$(\tilde{u}^0_{L},\tilde{u}^0_{R})$ and $(\tilde{u}^{0,*}_{L},\tilde{u}^{0,*}_{R})$, the mass squared matrix for up-squark is:
\begin{equation}
m^2_{\tilde{u}} = \left(
\begin{array}{cc}
m_{\tilde{u}_L^0\tilde{u}_L^{0,*}} &\frac{1}{\sqrt{2}}  v_u T_{u}^{\dagger}  -\frac{1}{2} v_d {\lambda} v_S Y_{u}^{\dagger} \\
\frac{1}{\sqrt{2}}  v_u T_{u}  - \frac{1}{2}v_d {\lambda}^* v_S Y_{u}  &m_{\tilde{u}_R^0\tilde{u}_R^{0,*}}\end{array}
\right),
\end{equation}
\begin{eqnarray}
&&m_{\tilde{u}_L^0\tilde{u}_L^{0,*}} = \frac{1}{24} \Big((g_1^2 +g_{Y B}^2-3g_2^2
+ g_{Y B} g_{B} )(v_{u}^{2}- v_{d}^{2})+ 2(g_{B}^2+ g_{Y B} g_{B})(v_{\bar{\eta}}^{2}-v_{\eta}^{2})
\Big)\nonumber \\&&\hspace{1.4cm}+\frac{1}{2} v_{u}^{2}Y_{u}^\dagger Y_{u}+m_{\tilde{Q}}^2 ,\nonumber\\&&
m_{\tilde{u}_R^0\tilde{u}_R^{0,*}} = \frac{1}{24}\Big((4g_{1}^{2}+4g_{Y B}^{2}+g_{Y B} g_{B})(v_{d}^{2}-v_{u}^{2})+2(4g_{Y B} g_{B}+g_{B}^2)(v_{\eta}^{2}-v_{\bar{\eta}}^{2})
\Big)\nonumber \\&&\hspace{1.4cm}+\frac{1}{2} v_{u}^{2}Y_{u} Y_{u}^\dagger+m_{\tilde{U}}^2.
\end{eqnarray}

This matrix is diagonalized by $Z^U$:
\begin{eqnarray}
Z^U m^2_{\tilde{u}}Z^{U,\dagger} = m^{dia}_{2,\tilde{u}}.
\end{eqnarray}

In the same way, we obtain the mass squared matrix for down-squark:
\begin{equation}
m^2_{\tilde{d}} = \left(
\begin{array}{cc}
m_{\tilde{d}_L^0\tilde{d}_L^{0,*}} &\frac{1}{\sqrt{2}}  v_d T_{d}^{\dagger}  -\frac{1}{2} v_u {\lambda} v_S Y_{d}^{\dagger} \\
\frac{1}{\sqrt{2}}  v_d T_{d}  - \frac{1}{2}v_u {\lambda}^* v_S Y_{d}  &m_{\tilde{d}_R^0\tilde{d}_R^{0,*}}\end{array}
\right),
\end{equation}
\begin{eqnarray}
&&m_{\tilde{d}_L^0\tilde{d}_L^{0,*}} = \frac{1}{24} \Big((g_1^2 +g_{Y B}^2+3g_2^2
+ g_{Y B} g_{B} )(v_{u}^{2}- v_{d}^{2})+ 2(g_{B}^2+ g_{Y B} g_{B})( v_{\bar{\eta}}^{2}-v_{\eta}^{2})
\Big)\nonumber \\&&\hspace{1.4cm}+\frac{1}{2} v_{d}^{2}Y_{d}^\dagger Y_{d} +m_{\tilde{Q}}^2,\nonumber\\&&
m_{\tilde{d}_R^0\tilde{d}_R^{0,*}} = \frac{1}{24}\Big((g_{Y B} g_{B}-2g_{1}^{2}-2g_{Y B}^{2})(v_{d}^{2}-v_{u}^{2})+2(g_{B}^{2}-2g_{Y B} g_{B})(v_{\eta}^{2}-v_{\bar{\eta}}^{2})
\Big)\nonumber \\&&\hspace{1.4cm}+\frac{1}{2} v_{d}^{2}Y_{d}Y_{d}^\dagger +m_{\tilde{D}}^2.
\end{eqnarray}

This matrix is diagonalized by $Z^D$:
\begin{eqnarray}
Z^D m^2_{\tilde{d}}Z^{D,\dagger} = m^{dia}_{2,\tilde{d}}.
\end{eqnarray}

To save space in the text, other mass matrixes can be found in Ref. \cite{han}.

We clarify certain couplings that are required for subsequent applications within the framework of this model. In the below equations, $P_L=\frac{1}{2}{(1 - {\gamma _5})}$, $P_R=\frac{1}{2}{(1 + {\gamma _5})}$.

1. The vertexes of $\bar{l}_i-\chi_j^--\tilde{\nu}^R_k(\tilde{\nu}^I_k)$
\begin{eqnarray}
&&\mathcal{L}_{\bar{l}\chi^-\tilde{\nu}^R}=\frac{1}{\sqrt{2}}\bar{l}_i\Big\{U^*_{j2}Z^{R*}_{ki}Y_l^iP_L
-g_2V_{j1}Z^{R*}_{ki}P_R\Big\}\chi_j^-\tilde{\nu}^R_k,\nonumber\\
&&\mathcal{L}_{\bar{l}\chi^-\tilde{\nu}^I}=\frac{i}{\sqrt{2}}\bar{l}_i\Big\{U^*_{j2}Z^{I*}_{ki}Y_l^iP_L
-g_2V_{j1}Z^{I*}_{ki}P_R\Big\}\chi_j^-\tilde{\nu}^I_k.
\end{eqnarray}

2. The vertexes of $\bar{\chi}_i^0-l_j-\tilde{L}_k$
\begin{eqnarray}
&&\mathcal{L}_{\bar{\chi}^0l\tilde{L}}=\bar{\chi}^0_i\Big\{\Big[\frac{1}{\sqrt{2}}\Big(g_1N^*_{i1}+g_2N^*_{i2}+(g_{YB}+g_B)N^*_{i5}\Big)Z^E_{kj}
-N^*_{i3}Y^j_lZ^E_{k(3+j)}\Big]P_L\nonumber\\&&\hspace{1.6cm}
-\Big[\frac{1}{\sqrt{2}}\Big(2g_1N_{i1}+(2g_{YB}+g_B)N_{i5}\Big)Z^E_{k(3+j)}+Y_{l}^jZ^E_{kj}N_{i3}\Big]P_R\Big\}l_j\tilde{L}_k.
\end{eqnarray}

3. The vertexes of $Z_{\mu}-{\chi}_i^{\pm}-{\chi}_j^{\pm}$
\begin{eqnarray}
&&\mathcal{L}_{Z {\chi}^{\pm} {\chi}^{\pm}}= \bar{\chi}^{\pm}_i\Big\{\frac{1}{2} \Big(2 g_{2} U_{j1}^{*} \cos\theta_{W} \cos\theta_{W}' U_{i1}+U_{j2}^{*} (-g_{1} \cos\theta_{W}' \sin\theta_{W} \nonumber\\&&\hspace{1.4cm}+g_{2} \cos\theta_{W} \cos\theta_{W}'+g_{Y B} \sin\theta_{W}') U_{i2}\Big) {\gamma}_{\mu}P_L\nonumber\\&&\hspace{1.4cm}+ \frac{1}{2} \Big(2 g_{2} V_{i1}^{*} \cos\theta_{W} \cos\theta_{W}' V_{j1}+V_{i2}^{*} (-g_{1} \cos\theta_{W}' \sin\theta_{W} \nonumber\\&&\hspace{1.4cm}+g_{2} \cos\theta_{W} \cos\theta_{W}'+g_{Y B} \sin\theta_{W}') V_{j2}\Big) {\gamma}_{\mu}P_R\Big\}{\chi}_j^{\pm}Z_{\mu}.
\end{eqnarray}

4. The vertex of $Z-\tilde{e}_i-\tilde{e}^{*}_j$
\begin{eqnarray}
&&\mathcal{L}_{Z\tilde{e}\tilde{e}^{*}}=\frac{1}{2}\tilde{e}^{*}_j
\Big[\Big(g_2\cos\theta_W\cos\theta_W^\prime-g_1\cos\theta_W^\prime\sin\theta_W
+(g_{YB}+g_B)\sin\theta_W^\prime\Big)\sum_{a=1}^3Z_{i,a}^{E,*}Z_{j,a}^E\nonumber\\&&\hspace{1.2cm}
+\Big((2g_{YB}+g_B)\sin\theta_W^\prime-2g_1\cos\theta_W^\prime\sin\theta_W\Big)
\sum_{a=1}^3Z_{i,3+a}^{E,*}Z_{j,3+a}^E\Big](p^\mu_{i}-p^\mu_j)\tilde{e}_iZ_{\mu}.
\end{eqnarray}

5. The quark-related vertices
\begin{eqnarray}
&&\mathcal{L}_{\chi^0d\tilde{D}}=-\frac{i}{6}\bar{\chi}^0_i\Big\{\Big[\sqrt{2}\Big(g_1 N_{1i}-3 g_2 N_{2i} + (g_{Y B}+g_B) N_{5i}\Big)Z^{\tilde{D}*}_{jk} +6 N_{3i} Y_d^j Z^{\tilde{D}*}_{(3+j)k} \Big]P_L\nonumber\\
&&\hspace{1.4cm}+\Big[6 Y_d^j Z^{\tilde{D}*}_{jk}  N^*_{3i}
+ \sqrt{2} Z^{\tilde{D}*}_{(3+j)k}\Big(2 g_1 N^*_{1i} + (2 g_{YB} - g_{B})N^*_{5i}\Big)\Big]P_R\Big\}d_j\tilde{D}^*_k,
\end{eqnarray}
\begin{eqnarray}
&&\mathcal{L}_{\chi^0u\tilde{U}}=-\frac{i}{6}\bar{\chi}^0_i\Big\{ \Big[\sqrt{2}\Big( g_1 N_{1i} +3  g_2 N_{2i} + (g_{YB}+g_B) N_{5i}\Big) Z^{\tilde{U}*}_{jk}+6 N_{4i} Y_u^j Z^{\tilde{U}*}_{(3+j)k} \Big]P_L\nonumber\\
&&\hspace{1.6cm}- \Big[ \sqrt{2}Z^{\tilde{U}*}_{(3+j)k}  \Big((g_{B} + 4g_{Y B})N^*_{5i} + 4 g_1 N^*_{1i}\Big) -6 Y_u^j Z^{\tilde{U}*}_{jk} N^*_{4i}\Big] P_R\Big\}u_j\tilde{U}^*_k,
\end{eqnarray}
\begin{eqnarray}
&&\mathcal{L}_{\chi^-d\tilde{U}}=\bar{d}_i\Big\{U^*_{j2}\sum_{a=1}^3 Z^{\tilde{U}*}_{ki} Y_d^i P_L +\Big[\sum_{a=1}^3 Y_u^i Z^{\tilde{U}*}_{k(3+i)}V_{j2}-g_2\sum_{a=1}^3 Z^{\tilde{U}*}_{ki}V_{j1}\Big]P_R\Big\}{\chi}^-_i\tilde{U}^*_k,
\end{eqnarray}
\begin{eqnarray}
&&\mathcal{L}_{\chi^-u\tilde{D}}=\bar{\chi}^-_i\Big\{\Big[U^*_{i2} \sum_{a=1}^3 Y_d^j Z^{\tilde{D}}_{k(3+j)}-g_2 U^*_{i1} \sum_{a=1}^3 Z^{\tilde{D}}_{kj}\Big] P_L+\sum_{a=1}^3 Z^{\tilde{D}}_{kj} Y_u^{j*} V_{i2} P_R\Big\}{u}_j\tilde{D}^*_k.
\end{eqnarray}


\begin{thebibliography}{50}
\vspace{3mm}
\bibitem{neutrino1}K. Abe et al., \emph{Phys. Rev. Lett.} {\bf107} (2011) 041801.
\bibitem{neutrino2}J. Ahn et al., \emph{Phys. Rev. Lett.} {\bf108} (2012) 191802.
\bibitem{neutrino3}F.An et al., \emph{Phys. Rev. Lett.} {\bf108} (2012) 171803.
\bibitem{neutrinoN1}E. Ma, A. Natale, O. Popov, \emph{Phys. Lett. B} {\bf746} (2015) 114-116.
\bibitem{neutrinoN2}I. Girardi , S.T. Petcov , A.V. Titov, \emph{Nucl. Phys. B} {\bf894} (2015) 733-768.
\bibitem{neutrinoN3}P. Ghosh, S. Roy, \emph{J. High Energy Phys.} {\bf0904} (2009) 069.
\bibitem{neutrinoN4}P. Ghosh, P. Dey, B. Mukhopadhyaya, S. Roy, \emph{J. High Energy Phys.} {\bf1005} (2010) 087.
\bibitem{p1}S. T. Petcov, Sov. \emph{J. Nucl. Phys.} {\bf 25} (1977) 340.
\bibitem{pdg}S. Navas et al., \emph{Phys. Rev. D} {\bf110} (2024) 3, 030001.
\bibitem{experiment1}P. Paradisi, \emph{J. High Energy Phys.}, {\bf10} (2005) 006.
\bibitem{experiment2}J. Girrbach, S. Mertens, U. Nierste and S. Wiesenfeldt, \emph{J. High Energy Phys.} {\bf05} (2010) 026.
\bibitem{experiment3}J. Rosiek, P. H. Chankowski, A. Dedes, S. Jager and P. Tanedo, \emph{Comput. Phys. Commun.} {\bf181} (2010) 2180.
\bibitem{mu}U. Ellwanger, C. Hugonie, A.M. Teixeira, \emph{Phys. Rep.} {\bf 496} (2010) 1-77.
\bibitem{neutrino4}B. Yan, S.M. Zhao, T.F. Feng, \emph{Nucl. Phys. B} {\bf 975} (2022) 115671.
\bibitem{han}X.Y. Han, S.M. Zhao, L. Ruan, et al., \emph{Eur. Phys. J. C} {\bf 85} (2025) 2, 163.
\bibitem{ZHB1}H.B. Zhang, T.F. Feng, L.N. Kou, et al., \emph{Int.J.Mod.Phys. A} {\bf 28} (2013) 24, 1350117.
\bibitem{ZHB2}H.B. Zhang, T.F. Feng, S.M. Zhao, et al., \emph{Nucl.Phys. B} {\bf 873} (2013) 300-324, Errutum: \emph{Nucl. Phys. B} {\bf879} (2014) 235.
\bibitem{nuRMSSM}A. Ilakovac, A. Pilaftsis, L. Popov, \emph{Phys. Rev. D} {\bf 87} (2013) 053014.
\bibitem{ljlig}J. Hisano, T. Moroi, K. Tobe, et al., \emph{Phys. Rev. D} {\bf 53} (1996) 2442.
\bibitem{BLMSSM3}P.F. Perez, M.B. Wise, \emph{J. High Energy Phys.} {\bf 1108} (2011) 068.
\bibitem{BLMSSM4}P.F. Perez, M.B. Wise, \emph{Phys. Rev. D} {\bf 82} (2010) 011901.
\bibitem{GT}T. Guo, S.M. Zhao, X.X. Dong, et al., \emph{Eur. Phys. J. C} {\bf 78} (2018) 11, 925.
\bibitem{LCTHiggs1}M. Carena, J.R. Espinosaos, C.E.M. Wagner, et al., \emph{Phys. Lett. B.} {\bf 355} (1995) 209.
\bibitem{LCTHiggs2}M. Carena, S. Gori, N.R. Shah, et al., \emph{J. High Energy Phys.} {\bf 1203} (2012) 014.
\bibitem{UMSSM5}G. Belanger, J.D. Silva, H.M. Tran, \emph{Phys. Rev. D} {\bf95} (2017) 115017.
\bibitem{B-L1}V. Barger, P.F. Perez, S. Spinner, \emph{Phys. Rev. Lett.} {\bf102} (2009) 181802.
\bibitem{B-L2}P.H. Chankowski, S. Pokorski, J. Wagner, \emph{Eur. Phys. J. C} {\bf47} (2006) 187.
\bibitem{gaugemass}J.L. Yang, T.F. Feng, S.M. Zhao, et al., \emph{Eur. Phys. J. C} {\bf78} (2018) 714.
\bibitem{Bernabeu:1993ta}J. Bernabeu, E. Nardi, D. Tommasini, \emph{Nucl. Phys. B} {\bf409} (1993) 69-86.
\bibitem{Sens:1959zz}J.C. Sens, \emph{Phys. Rev.} {\bf113} (1959), 679-687.
\bibitem{Zeff2}H.C. Chiang, E. Oset, T.S. Kosmas, et al., \emph{Nucl. Phys. A} {\bf559} (1993) 526.
\bibitem{Kitano:2002mt}R. Kitano, M. Koike, Y. Okada, \emph{Phys. Rev. D} {\bf66} (2002) 096002.
\bibitem{cms}CMS collaboration, \emph{Phys. Lett. B} {\bf716} (2012) 30.
\bibitem{atlas}ATLAS collaboration, \emph{Phys. Lett. B} {\bf716} (2012) 1.
\bibitem{w1}P. Cox, C.C. Han, and T.T. Yanagida, \emph{Phys. Rev. D} {\bf 104} (2021) 075035.
\bibitem{w2}M.V. Beekveld, W. Beenakker, M. Schutten, et al., \emph{SciPost Phys.} {\bf11} (2021) 3, 049.
\bibitem{w3} M. Chakraborti, L. Roszkowski and S. Trojanowski, \emph{J. High Energy Phys.} {\bf 05} (2021) 252.
\bibitem{w4}F. Wang, L. Wu, Y. Xiao, et al., \emph{Nucl. Phys. B} {\bf 970} (2021) 115486.
\bibitem{w5}M. Chakraborti, S. Heinemeyer and I. Saha, \emph{Eur. Phys. J. C} {\bf81} (2021) 12, 1114.
\bibitem{w6}M. Endo, K. Hamaguchi, S. Iwamoto, et al., \emph{J. High Energy Phys.} {\bf 07} (2021) 075.
\bibitem{lb}L. Basso, \emph{Adv. High Energy Phys.} {\bf 2015} (2015) 980687.
\bibitem{at}ATLAS collaboration, \emph{Phys. Lett. B} {\bf 796} (2019) 68.
\bibitem{gc}G. Cacciapaglia, C. Cs$\acute{a}$ki, G. Marandella, et al., \emph{Phys. Rev. D} {\bf 74} (2006) 033011.
\bibitem{ZPG2} M. Carena, A. Daleo, B. A. Dobrescu, et al., \emph{Phys. Rev. D} {\bf70} (2004) 093009.
\bibitem{HAN1} M. Drees, M. Gluck and K. Grassie, \emph{Phys. Lett. B} {\bf157} (1985) 164-168.
\bibitem{HAN2} U. Chattopadhyay, D. Das and S. Mukherjee, \emph{J. High Energy Phys.} {\bf06} (2020) 015.
\bibitem{Zhao:2015dna}S.M. Zhao, T.F. Feng, H.B. Zhang, et al., \emph{J. High Energy Phys.} {\bf92} (2015) 115016.
\bibitem{L1}S.M. Zhao, L.H. Su, X.X. Dong, et al., \emph{J. High Energy Phys.} {\bf03} (2022) 101.
\bibitem{L2} Muon g-2 collaboration, \emph{Phys. Rev. D} {\bf 73} (2006) 072003.
\bibitem{L3}Muon g-2 collaboration, \emph{Phys. Rev. Lett.} {\bf 126} (2021) 141801.


\end{thebibliography}
\end{document}